\documentclass[aps, prd, 10pt, superscriptaddress, showpacs, twocolumn,floatfix]{revtex4-2}
\usepackage{amsmath,amsfonts,amssymb,graphicx,color, bm, xcolor, units}
\usepackage[colorlinks=true,citecolor=blue]{hyperref}

\newcommand{\de}{\partial}

\newcommand{\be}{\begin{equation}}
\newcommand{\ee}{\end{equation}}
\DeclareMathOperator{\Tr}{Tr}

\begin{document}

\title{Minimal superfluid vortices in chiral
perturbation theory }
\author{Fabrizio Canfora}
\email{fabrizio.canfora@uss.cl}
\affiliation{Centro de Estudios Cient\'{\i}ficos (CECS), Casilla 1469, Valdivia, Chile}
\affiliation{Universidad San Sebasti\'{a}n, sede Valdivia, General Lagos 1163,
Valdivia 5110693, Chile}

\author{Mart\'{\i}n Garrido}
\email{margarrido2021@udec.cl}
\affiliation{Universidad de Concepci\'on (UDEC), Concepci\'on, Chile}

\author{Massimo Mannarelli}
\email{massimo.mannarelli@lngs.infn.it}
\affiliation{Laboratori Nazionali del Gran Sasso, INFN, Via G.~Acitelli 22,
Assergi (AQ), I-67100, Italy}

\author{Anibal Neira}
\email{aneira2017@udec.cl}
\affiliation{Universidad de Concepci\'on (UDEC), Concepci\'on, Chile}

\begin{abstract}
We derive some properties of rotational vortices in the pion condensed phase. Employing leading order chiral perturbation theory we determine the minimal energy condition for vortex nucleation. Vortices have quantized angular momentum along the rotation axis, a hallmark of superfluidity, and self-confined pions. The critical rotation frequency for vortex nucleation is estimated.
\end{abstract}

\maketitle

\section{Introduction} 

\label{sec-1}

The theoretical understanding  of the phases  of strongly interacting matter  at different temperature, baryonic density and isospin asymmetry is one of the major challenges in particle physics.
Many of the most
important open problems in quantum-chromo dynamics (QCD)  are indeed deeply linked to the possible realization of the different phases of hadronic matter. For instance, the mechanism underlying  color confinement can be explored studying the high-temperature transition between standard hadronic matter and the quark-gluon plasma (QGP)\,\cite{Shuryak:2014zxa, Pasechnik:2016wkt,Busza:2018rrf, Bzdak:2019pkr, Yagi, Kogut:2004su, Brambilla:2014jmp, Pisarski:2022abl},
or  determining whether a  low-temperature  high-density    transition to   quarkyonic\,\cite{McLerran:2007qj, *Hidaka:2008yy} or color superconducting\,\cite{Alford:1997zt, Rajagopal:2000wf, Alford:2007xm, Anglani:2013gfu} matter occurs.   Unfortunately,  these   phase transitions  cannot be attacked by perturbative methods:  at the   relevant energy  scales QCD is nonperturbative. Therefore, alternative methods must be used.

Numerical  lattice QCD (LQCD) simulations allow one to explore a sizable part of the  hadronic matter phase diagram\,\cite{Aoki:2006we, Bhattacharya:2014ara}.   However,  the so-called sign problem  limits  the  region of  parameter space where this method can be used\,\cite{Bzdak:2019pkr, Nagata:2021ugx, Astrakhantsev:2021fua, Astrakhantsev:2021uii, Brandt:2023dir, Busza:2018rrf, Yagi:2005yb}. In particular, investigating   hadronic matter as a function of the baryonic chemical potential is challenging\,\cite{deForcrand:2009zkb}. This means that the quarkyonic phase as well as the color superconducting phase remain, so far,  inaccessible.  On the other hand, 
there is no numerical problem in  Monte Carlo simulations at vanishing baryonic density and nonzero isospin chemical potential, $\mu_I$\,\cite{Alford:1998sd}. Such system can be viewed as made of strongly interacting pions, which are stable because the weak interaction has been turned off.   At sufficiently low temperature it happens that for isospin chemical potential exceeding the pion mass, $m_\pi$, pions form an homogeneous  Bose-Einstein condensate (BEC). This phase  
was already proposed in the 1970s~\cite{Migdal:1971cu,Migdal_1972,Sawyer:1972cq, Scalapino:1972fu,1972PhLA...41..129K}  as the ground state of  extremely high-density nuclear matter. Although  its possible realization  is still controversial, see for instance\,\cite{Mannarelli:2019hgn}, nevertheless it represents a formidable portal to inquiry the properties of hadronic matter. One of the reasons is that it can be studied by means of several somehow complementary  methods: Chiral perturbation theory ($\chi$PT)\cite{Baym:1978sz, Kaplan:1986yq, Dominguez:1993kr, Son:2000xc, Kogut:2001id, Birse:2001sn, Splittorff:2002xn, Loewe:2002tw, Loewe:2004mu, Loewe:2011tm, Mammarella:2015pxa, Carignano:2016rvs, Loewe:2016wsk, Carignano:2016lxe, Lepori:2019vec, Adhikari:2019mdk, Tawfik:2019tkp, Mishustin:2019otg},  NJL models\,\cite{Barducci:1990sv,Toublan:2003tt, Barducci:2004tt,Barducci:2004nc, He:2005nk, Ebert:2005cs, Ebert:2005wr, Mukherjee:2006hq, He:2005sp, He:2006tn, Sun:2007fc, Andersen:2007qv,Abuki:2008tx, Abuki:2008wm, Mu:2010zz,Xia:2013caa,  Xia:2014bla, Chao:2018ejd, Khunjua:2018jmn,Khunjua:2019lbv, Khunjua:2019nnv,Avancini:2019ego,Lu:2019diy,Carlomagno:2024xmi},  linear $\sigma$ and quark-meson models\,\cite{Klevansky:1992qe, Andersen:2014xxa, Adhikari:2016eef, Adhikari:2018cea,Andersen:2018osr,Andersen:2018qkq,Brandt:2025tkg}, random matrix models\,\cite{Klein:2003fy,Klein:2004hv}, ladder QCD\,\cite{Barducci:2003un}  and holographic QCD \cite{Lv:2018wfq,Kovensky:2021ddl, Chen:2024cxh} allow one to understand many  features of meson condensation. The findings of  effective theories align with   LQCD results~\cite{ Kogut:2002tm, Kogut:2002zg,Kogut:2004zg, Beane:2007es, deForcrand:2007uz, Detmold:2008fn,Detmold:2008yn,Detmold:2011kw, Detmold:2012wc, Cea:2012ev, Endrodi:2014lja, Janssen:2015lda, Brandt:2016zdy, Brandt:2018omg, Brandt:2017zck, Brandt:2018wkp, Brandt:2022hwy, Brandt:2025ywy}  for $\mu_I \gtrsim m_\pi$, and this fact makes our description of the pion condensed phase robust.    
In particular, it is well established that there exists a second order phase transition between the naive vacuum and the pion condensed phase~\cite{Son:2000xc,Kogut:2001id}. In the range $\mu_I \gg m_\pi$, the errors in  LQCD simulations become large,  however they show  qualitative agreement with the outcomes of perturbative methods\,\cite{Graf:2015pyl, Andersen:2015eoa, Danhoni:2025qpn}.

One of the benefits of  effective theories is that they allow one to    handle the properties  of topological objects. These are of great importance because   they encode valuable information on the non-perturbative sector of a theory\,\cite{Manton:2004tk, Shifman:2012zz, Shuryak:2021vnj, Kogut:2004su, Brambilla:2014jmp, Pisarski:2022abl,Weinberg:2012pjx}.
A prototypical example is the case of vortices in type II  superconductors \cite{Manton:2004tk,Weinberg:2012pjx}. 
Superconducting vortices with quantized magnetic flux present in the mixed phase characterize the  response of these systems to an external magnetic field. However, the theoretical description of type II  superconductors is challenging as the actual mechanism underlying  superconductivity is poorly understood.  The simplest theory, the one elaborated by  Bardeen, Cooper and  Schrieffer (BCS), can qualitatively describe various properties of type I superconductors, but struggles with type II superconductors, which are typically realized with complex materials.

A powerful technique to study such topological configurations was  provided by Bogomol'nyi, Prasad and Sommerfield (BPS), that obtained  the minimum energy of soliton configurations, known as the BPS bound. In this case, the actual expression of interactions becomes immaterial, insofar they allow the realization of the broken phase. This happens at a particular value of the system parameters, called the BPS critical point. 
In case of superconducting vortices it corresponds to the
critical point between
Type II and Type I superconductors. In the Ginzburg–Landau theory, the  penetration lengths of the   scalar and vector gauge fields are equal, the ground state is then determined by the solution of  first order nonlinear differential equations, instead of  the
typical second order field equations. A further advantage to
deal with solitons at the BPS critical point is that this condition allows us to study the low-energy dynamics using the so-called
moduli-space approximation, see 
\cite{Manton:1981mp, *Manton:1985hs, Manton:1998kq, Gibbons:1995yw}.
Moreover, the analysis of interaction of 
solitons with fermions is simplified by  powerful index theorems\,\cite{Manton:2004tk,Weinberg:2012pjx}. Thus, the theory of BPS solitons is not just a mathematical
curiosity but rather a very efficient tool for the analysis of nonperturbative states.  It can  be applied to standard superfluids  described by the
Gross-Pitaevskii (GP) equation\,\cite{Gross1961soa, Pitaevskii1961vli, Dalfovo1999tob, Schmitt:2014eka, Stringari2016bec}. A BPS bound was  indeed derived in two-dimensional GP theory\,\cite{Canfora:2025jqr,Canfora:2025qkl} together with the corresponding first order BPS differential equations. These results  suggest that a similar construction could work   in $\chi $PT\,\cite{COMTET1988719,Canfora:2026col}.  Using $\chi $PT minimally coupled to electromagnetism  has been possible to find the BPS bound in two different systems of pions\,\cite{Canfora:2023zmt,Canfora:2024mkp}.

In the present paper we  employ leading order $\chi$PT at the critical BPS point to  
show that the pion  condensed phase can host quantized vortices. Vortices arise as topological solutions of the effective field theory Lagrangian and in this context it turns out that they become stable above a critical rotation frequency. Since the vortex solutions that we study  are   found combining  leading order $\chi$PT  with the BPS critical condition, they inherit the same restrictions on the parameter space needed  for the validity of these two theories. The system should be close to ground state, thus we assume vanishing temperature,  and the isospin chemical potential should be close to the pion mass, otherwise $\chi$PT breaks down.  In addition, the realization of the BPS condition requires that the isospin chemical potential should be appropriately tuned.  
Considering    cylindrical Minkowskian  space-time at vanishing temperature,  vortex solutions with minimal free-energy cost are found   by  appropriately tuning  the isospin chemical potential.
Remarkably,  the results we obtain show close analogy with those long studied in standard superfluids and in dilute  ultracold atoms: the total angular momentum  is quantized and vortex nucleation happens at a critical frequency depending on the system size. The main -- and   quite interesting --  different feature is that  matter confinement is not determined by an external vessel or a trap, it is instead self-induced by the pion interaction.

This paper is organized as follows. In Sec.\,\ref{sec:setup} we introduce the 
$\chi$PT theoretical setup and we briefly 
recap the main features of the  well-known homogeneous pion condensed phase.
In Sec.\,\ref{sec:inhomogeneous} we derive the BPS bound in $\chi $PT at nonvanishing isospin chemical potential. In Sec.\, \ref{sec:single} we construct superfluid single vortex solutions at the BPS critical point. More general vortex structures are discussed in Sec.\,\ref{sec:general_vortex}, while the free energy and the critical rotation frequency are studied in Sec.\,\ref{sec:free-energy}. We draw our conclusions in Sec.\,\ref{sec:conclusion}.
In the Appendix we  provide some details of the calculation of the 
angular momentum of noncentral vortices. We employ natural units  $c=%
\hslash =1$.


\section{Theoretical setup}

\label{sec:setup} 
A system of pions at vanishing temperature in  Minkowski spacetime $\mathcal{M}$ can be described by the  $\chi $PT  
action
\be\label{eq:action}
S=\int_{\mathcal{M}}\!\!d^{4}x\left( K\Tr\left[
L^{\mu }L_{\mu }^\dag \right] +V\right) \,, 
\ee
which only includes the leading   $\mathit{O}(p^{2})$  terms and it is valid for momenta $p\ll \Lambda_\chi$, where $\Lambda_\chi \sim 1 $ GeV is the $\chi$PT breaking scale, see \,\cite{Weinberg:1978kz, Gasser:1983yg, Georgi:1985kw, Leutwyler:1993iq,
Ecker:1994gg, Leutwyler:1996er, Pich:1998xt, Scherer:2002tk, Scherer:2005ri}. Here, $K=f_{\pi}^2/4$, with $f_{\pi }$ the pion decay constant and  $V$  the potential that will be specified below.  The derivative term has been conveniently written by means of  
\be
L_{\mu }=U^{-1}{D}_{\mu }U\,,
\ee
where $U(x)\in SU(2)$ is an isospin  representation of the  pion fields,
while the covariant derivative 
\begin{equation}
{D}_{\mu }U=\de _{\mu }U+i\frac{\mu _{I}}{2}[\sigma_{3},U]\delta_{\mu 0}\, ,  \label{muI}
\end{equation}%
includes the contribution of the isospin chemical potential. This can be viewed as a time-like  external vector field proportional to the Pauli matrix $\sigma_3$. It explicitly breaks boost invariance, because it is proportional to $\delta_{\mu 0}$, and   the $ SU(2)$ isospin symmetry:   the  Lagrangian is solely invariant under transformations of the $U(1)$ subgroup generated by $\sigma_3$; see\,\cite{Mannarelli:2019hgn} for more details.   

By comparison with LQCD results, one finds that  Eq.\,\eqref{eq:action} allows us to  accurately  describe the low-energy properties of pions for $\mu_{I}\lesssim 1.2m_{\pi }$;  higher values of the isospin chemical potential require  additional terms in the action, see for instance\,\cite{Kojo:2021hqh}. In the  representation 
\be\label{eq:U}
U=\boldsymbol{1} \cos \alpha +i \bm n \cdot \bm \sigma \sin \alpha \,,  \ee
where $
\boldsymbol{1}$ denotes the $2\times 2$ identity matrix, $\bm\sigma_{i}$ for $i=1, 2, 3 $  are the Pauli matrices and pions are grouped in the unit vector  
\be\label{eq:n}
\bm n=\{\sin \Theta \cos \Phi ,\sin \Theta \sin \Phi ,\cos \Theta \}\ , 
\ee
where  $\alpha =\alpha (x^{\mu })$, $\Theta =\Theta (x^{\mu })$, $\Phi =\Phi
(x^{\mu })$  are   three real scalar fields. In order to make contact with the standard representation of pions,  it is useful to compare  Eqs.\,\eqref{eq:U} and \eqref{eq:n} with
\be\label{eq:pion}
U = \exp\left(i \frac{\bm \sigma \cdot \bm \varphi}{f_\pi}\right)\,,
\ee
where $\bm \varphi = (\varphi_1,\varphi_2,\varphi_3)$ are three scalar fields and  pions  correspond to
\be
\pi_0 = \varphi_3 \qquad \pi_\pm = \frac{ \varphi_1 \pm i \varphi_2}{\sqrt{2}}\,.
\ee
Matching the two representations   we have that   $\pi_0=\varphi_3 = \alpha f_\pi \cos\Theta$ and thus  a system without the condensation of neutral pions corresponds to  $\Theta = \pi/2$. Since neutral pions are  states with $I_3 =0$, at  tree level they  are insensitive to  the isospin chemical potential. In the following we will  be interested in the  phases that appear as a response to a nonvanishing isospin chemical potential, thus  taking $\varphi_3 = 0$ 
is a reasonable assumption; as will be shown hereafter, it corresponds to a minimum of the free energy. For vanishing neutral pion component,  we have that 
\be
\alpha\, e^{\pm i \Phi} = 
\sqrt{2} \frac{\pi_{\pm}}{f_\pi}\,,
\ee
and therefore
\be\label{eq:alpha_phi}
\alpha = \frac{\sqrt{2}}{f_\pi} \pi_+ \pi_- \quad \text{and} \quad \tan \Phi = i\frac{\pi_+ + \pi_-}{\pi_- - \pi_+}\,,
\ee
meaning that  $\alpha$ describes the absolute value of the pion condensate, while  $\Phi$ is its phase.

Although Eq.\,\eqref{eq:pion} is useful to describe pion interactions,  the  ground state properties are more easily studied  using Eq.\,\eqref{eq:U}. With such parametrization, the Lagrangian density in Eq.\,\eqref{eq:action} can be written as
\be
{\mathcal L} = {\mathcal L}_k + \mathcal{L}_{V}\,,
\ee
with the derivative term  given by
\begin{align}
    {\mathcal L}_k = &
2 K \left[ \de_\mu \alpha \de^\mu \alpha + \sin^2\!\alpha  \de_\mu \Theta \de^\mu \Theta \right. \nonumber \\ & \left.+\sin^2\!\alpha \sin^2\!\Theta \left(\de_\mu \Phi \de^\mu \Phi- 2 \mu_I \delta_{\mu 0} \de^\mu \Phi\right) \right]\,,
\label{eq:LK}
\end{align}
which includes nonlinear couplings between the three fields. From Eq.\,\eqref{eq:alpha_phi} it is clear that the expansion of the  term proportional to $\sin^2\alpha$ produces   an infinite number of charged pion self-interactions.  The second Lagrangian contribution, $\mathcal{L}_{V}$, is related to the breaking of  chiral symmetry. At a fundamental level it depends on quark interactions and masses. While 
${\mathcal L}_k$ is the only independent $\mathit{O}(p^{2})$   Lagrangian   that respects  chiral symmetry,  for the potential term there is no such restriction.  
In the following we will employ 
\begin{align}
{\cal L}_V &= 2 K \mu_I^2 \sin^2\!\alpha \sin^2\!\Theta +  V \,,\label{eq:LV}
\end{align}
where the first contribution on the rhs comes from the expansion of the first term in the round bracket of Eq.\,\eqref{eq:action}, while
\be V=
K m_\pi^2  \Tr\left(U + U^\dag\right) = 4 K m_\pi^2 \cos\alpha\,,
\ee
is the standard $\chi$PT term giving masses to  pions; an expression that can be immediately generalized to $SU(3)$ symmetry. 
From the above Lagrangian, one can obtain the  free-energy density 
\be
F = F_K +  F_V\,,
\ee
where
\be
F_K= 2 K\left[\left( {\bm \nabla}\alpha
\right) ^{2}+\sin ^{2}\!\alpha  \left( {\bm \nabla}\Theta
\right) ^{2} +\sin ^{2}\!\alpha  \sin ^{2}\!\Theta \left( {\bm \nabla}\Phi \right)^2  \right]\,,
\label{eq:Fk}
\ee
contains the space derivatives, while
\begin{align}
F_V &=-2 K\mu_{I}^{2}\sin^{2}\!\alpha\,  \sin ^{2}\!\Theta   + 4K m_\pi^2 (1-\cos\alpha)\, ,  \label{eq:FV}
\end{align}
is the potential term;  here we have appropriately added the vacuum free energy, $4K m_\pi^2$, so that $F_V $ vanishes in the unbroken phase.

\subsection{Homogeneous phase}
\label{sec:homogeneous}
We briefly recall  the main properties of  the homogeneous and static pion condensed phase in $\chi$PT.  In this  case $\alpha, \Phi$  and $\Theta$ are treated as variational parameters determined by  minimizing the potential. Actually, $F_{V}$ is independent of $\Phi$, therefore this parameter cannot be fixed, while  $\Theta = \pi/2$  minimizes $F_V$ for any nonvanishing value of $\mu_I$ and $\alpha$.  As we have seen, this corresponds to a vanishing neutral pion component. {Any isospin rotation  $\Theta = \pi/2 + \delta\Theta$  would have a free-energy cost
\be
\Delta F_V =   2 K \mu_I^2 \sin^2\! \alpha \sin^2 \!\delta\Theta\,,
\ee
and it is therefore energetically forbidden. The physical reason is that
such rotation would imply that neutral pion states are populated, with a free-energy cost proportional to $\pi_0^2 = f_\pi^2 \cos^2\Theta=f_\pi^2 \sin^2\delta\Theta$ and the squared pion mass $m_\pi^2 \sim \mu_I^2$.}

Upon substituting  $\Theta = \pi/2$ in   Eq.\,\eqref{eq:FV} we have that $F_V =-2 K\mu_{I}^{2}\sin ^{2}\!\alpha  +4K m_\pi^2(1- \cos\alpha) $ and then the free-energy minimum corresponds to
\be
\alpha = \begin{cases} 0  & \text {for  } |\mu_I| < m_\pi \\ 
\arccos\left({\frac{m_\pi^2}{\mu_I^2}}\right) & \text {for  } |\mu_I| \geq m_\pi\,,
\end{cases}
\label{eq:alpha_hom}
\ee
therefore at $|\mu_I| = m_\pi$ the $U(1)$ subgroup generated by $\sigma_3$  is spontaneously broken and  there is a second order phase transition
from the naive vacuum  to a BEC of charged pions. Since in the broken phase the free-energy minimum is attained for $\cos\alpha = m_\pi^2/\mu_I^2$, it follows that  any phase can be described taking $\alpha \in [0,\pi/2)$. For definiteness, hereafter  we  assume that $\mu_I \geq 0$.

Pressure and isospin number density in the broken phase are respectively given by
\begin{align}
    P &= 
\frac{ 2 K}{ \mu_I^2} (\mu_I^2- m_\pi^2)^2 \quad \text{and} \quad
n_I  =\frac{4 K}{ \mu_I^3} (\mu_I^4 -  m_\pi^4)\,,
\end{align}
where the pressure is clearly isotropic. 
Using the
Gibbs--Duhem relation
$\displaystyle{ \epsilon = \mu n -P}$,
 we  obtain the  energy density 
\be
\epsilon = \frac{ 2 K }{ \mu_I^2} (\mu_I^2- m_\pi^2)(\mu_I^2+ 3 m_\pi^2)\,,
\ee
and then   the adiabatic speed of sound\,\cite{Carignano:2016lxe} by means of
\be
c_s = \sqrt{\frac{\de P}{\de \epsilon}}=  \sqrt{\frac{\mu_I^4 -m_\pi^4}{\mu_I^4 + 3m_\pi^4}}\,,
\ee
which is manifestly  homogeneous and isotropic.

\section{Inhomogeneous phases}
\label{sec:inhomogeneous}
We now move to inhomogeneous pion systems using as guidance the information gathered in  the previous analysis. As we have seen,  the homogeneous phase  free-energy minimum is always attained when $\Theta = \pi/2$, corresponding to a vanishing neutral pion component. On the other hand,  by changing the ``control parameter" $\mu_I$,   the value of $ \alpha$ that minimizes the free energy changes according to Eq.\,\eqref{eq:alpha_hom}. Such variation, as a function of $\mu_I$, is continuous.   It  thus appears natural to fix  $\Theta = \pi/2$ and then look for possible inhomogeneous phases that could be realized by a space modulation of $\alpha$.  The free energy of the homogeneous ground state does not depend on $\Phi$. The reason is that this field spans the flat direction of the potential and indeed  the associated fluctuation is   the Nambu-Goldstone boson. In general, space gradients of the phase  are associated to a velocity field, for this reason we will assume that $  \Phi$ is a space dependent field. This will allow us to consider vortices, characterized by a swirling velocity field.

To recap,   we will assume that $\Theta=\pi/2$ and that both $\alpha$ and $\Phi$ are classical space-dependent fields. 
In this case, the Lagrangian, see Eqs.\,\eqref{eq:LK} and \eqref{eq:LV}, turns into 
 \be\label{eq:L_sempl}
{\mathcal L} =
2 K \left[ \de_\mu \alpha \de^\mu \alpha + \sin^2\!\alpha  \de_\mu \Phi \de^\mu \Phi + 2 m_\pi^2 (\cos\alpha-1)\right]\,,
\ee
where we  found convenient to replace $\Phi \to \Phi + \mu_I t $
to have a more compact expression. This effective Lagrangian represents the starting point of our analysis. In particular, the vortex solutions we will analyze are derived starting from this Lagrangian. The free-energy contributions now read
\begin{align} F_K & = 2 K\left[\left( {\bm \nabla}\alpha
\right) ^{2}+\sin ^{2}\!\alpha  \left( {\bm \nabla}\Phi \right)^2  \right]\,, \label{eq:Fk_2}\\
F_V &=2 K\left[-\mu_{I}^{2}\sin ^{2}\!\alpha    + 2 m_\pi^2 (1-\cos\alpha)\right]\, ,\label{eq:FV_2}
\end{align}
and since we are interested in vortices, the natural ansatz is 
\begin{equation}
\alpha =\alpha (x_{1},x_{2})\ ,\quad \Phi =\Phi (x_{1},x_{2}) \,,\label{U1}
\end{equation}%
where the indices $1,2$ indicate the two coordinates in the $xy$-plane;  the  vorticity is then expected to be along the $z$-direction.

We can now rewrite the gradients  in Eq.\,\eqref{eq:Fk_2} as follows:
\begin{widetext}
\begin{align}
(\de_{1}\alpha )^{2}+(\de
_{2}\alpha )^{2}+\sin ^{2}\!\alpha [(\de _{1}\Phi )^{2}+(\de
_{2}\Phi )^{2}]=(\de _{1}\alpha \pm \sin\alpha\, \de
_{2}\Phi )^{2}+(\de _{2}\alpha \mp \sin\alpha\, \de _{1}\Phi )^{2}  &\pm 2\sin \alpha \left(  \de _{1}\Phi  \de
_{2}\alpha  -\de_{2}\Phi   \de
_{1}\alpha  \right) \,, 
\label{eq:alpha_dec}
\end{align}%
\end{widetext}
where the last term on the rhs of the above expression has the following property:
\begin{equation*}
\pm 2\sin (\alpha )\left[ \left( \de _{1}\Phi \right) \left( \de
_{2}\alpha \right) -\left( \de _{2}\Phi \right) \left( \de
_{1}\alpha \right) \right]dx_1dx_2 =d\omega \ ,
\end{equation*}%
with%
\begin{equation}
\omega =\pm 2\cos \alpha d\Phi \,,  \label{eq:dressedvorticity1}
\end{equation}%
 a $1$-form. 
If $\cos \alpha =1$,   then $\omega$ can be interpreted as the  vorticity describing the spinning of the $\Phi$ field.  Although in this case $d \omega =0$, the circulation does not  vanish if vortices are present.  Similarly to what happens  in the case of superconducting
vortices in $\chi $PT, see \cite{Canfora:2024mkp}, when $\cos \alpha \neq 1$, the obvious topological charge, $d \Phi$,
is dressed by the
background  modulation. In the following we will take the positive sign in Eq.\,\eqref{eq:dressedvorticity1}, and  consider solutions with $\cos\alpha >0$. 

Upon substituting Eqs.\,\eqref{eq:alpha_dec} and \eqref{eq:dressedvorticity1} in the  free-energy density in Eq.\,\eqref{eq:Fk_2}, we have that 
the total free energy satisfies the inequality
\be\label{eq:F_total}
{\cal F} = \int_M dx_1dx_2 F = {\cal F}_V + {\cal F}_K   \geq 2 K \int_M d\omega = 2 K \int_{\partial M} \omega\,,
\ee
and therefore, when  the free-energy potential vanishes, that is when
 \be\label{eq:BPS_condition}
\int_M d^3 x \left[-\mu_{I}^{2}\sin ^{2}\!\alpha  + 2 m_\pi^2 (1-\cos\alpha)\right]=0\,,
\ee
the total free energy equates ${\cal F}_K$. We will refer to this 
as the BPS condition. In addition, if 
the  BPS equations 
\begin{align}
\de_{1}\alpha \pm \sin\alpha\, \de _{2}\Phi =&0\,,
\label{eq:SFBPS1} \\
\de _{2}\alpha \mp \sin\alpha\,\de _{1}\Phi =&0\,,
\label{eq:SFBPS2}
\end{align}
hold, the inequality in \eqref{eq:F_total} is saturated, meaning that  
\be\label{eq:F_total_2}
{\cal F} =  2 K \int_{\partial M} \omega\,,
\ee
therefore the free energy is completely determined by the behavior of  vorticity at the boundary. This suggests, as we will see in the following, that it only depends  on the boundary of the manifold and on the total winding number. Since the free energy is positive,  vortex nucleation is not energetically favored. This happens because the system is not rotating, therefore 
there is no reason why a vortex should be nucleated. We will discuss the effect of rotation  in Sec.\,\ref{sec:free-energy}.   In numerical simulations with ultracold atoms adding a vortex to a static superfluid is usually obtained by the phase imprinting method. This procedure allows one to scrutinize, among the possible vortex configurations, the one with the lowest free-energy cost.

We will explore such configurations  employing the
BPS condition, Eq.\,\eqref{eq:BPS_condition}, which depends on the field $\alpha$, the isospin chemical potential and the system size. 
On the other hand, the BPS equations \eqref{eq:SFBPS1} and \eqref{eq:SFBPS2} involve $\alpha$ and $\Phi$. The set of equations \eqref{eq:BPS_condition}, \eqref{eq:SFBPS1} and \eqref{eq:SFBPS2} can be solved using the following procedure. First, we find the solutions of the differential equations \eqref{eq:SFBPS1} and \eqref{eq:SFBPS2} using appropriate boundary conditions. Then, we  determine  the constraint on  $\mu_I$ and  the system manifold to satisfy Eq.\,\eqref{eq:BPS_condition}.

\subsection{Boundary conditions and vorticity quantization}

For definiteness  we assume that the system is in  a cylinder  of radius $R$ and height $\ell_z$. We consider flat space-time and we employ  cylindrical coordinates
\begin{equation*}
ds^{2}=-dt^{2}+dr^{2}+r^{2}d\varphi^{2}+dz ^{2}\,,
\end{equation*}%
with $r$ and $\varphi$ the polar coordinates,  while $z$ is along the axis of the cylinder. The chiral field $U$ in Eq. \eqref{eq:U} must be single valued, therefore
\begin{equation}
U(r,\varphi )=U(r,\varphi  +2\pi )\,,  \label{singlevalued}
\end{equation}%
and  taking into account Eq. \eqref{eq:n}, with $\Theta=\pi/2$, we have that  for any $r$, 
\begin{align}\label{eq:multivalued_alpha}
\alpha (r,\varphi )&=\alpha (r,\varphi +2\pi )+2m\pi \,,\ \ m\in \mathbb{Z}\\
\Phi (r,\varphi )&=\Phi (r,\varphi +2\pi )+2n\pi \,, \ \ n\in \mathbb{Z}\,
\label{eq:multivalued_Phi}
\end{align}%
which implies that these fields are periodic functions of $\varphi$ or are linearly dependent on it with integer slope. In configurations that are cylindrically symmetric only the latter is possible, indeed a modulation along the tangential direction would break the rotational invariance around the $z$-axis.

In presence of vortices, the condition in Eq.\,\eqref{eq:multivalued_Phi} with $n \neq 0$ appears
natural taking into account that $\Phi $ is a phase. 
It implies that 
\begin{equation}
\frac{1}{2\pi }\oint_{\Gamma _{\infty }} \bm \nabla \Phi \cdot d\bm \ell=n\,,  \label{fluxquantization}
\end{equation}%
where $\Gamma _{\infty }$ is the circle at spatial infinity and $n$ is then the total winding number. 
If the cylindrical symmetry is broken, for instance by vortices not passing through the system center, then $\Phi$ becomes a combination of radial dependent functions and of sinusoidal  functions of $\varphi$ describing the winding  around each vortex. Regarding the field $\alpha$, it does not describe the phase of the condensate, but the  modulation of its absolute value, see
Eq.\,\eqref{eq:alpha_phi}. In cylindrically symmetric configurations we expect that
 $\alpha$ describes the radial modulation of the condensate induced by the  vortex.  For noncentral vortices $\alpha$ will be as well a sinusoidal function of $\varphi$. 

{Regarding the vortex topological stability, given the translational invariance along the $z$-direction, the system is effectively in $2+1$ dimensional spacetime with a $S^1$ space boundary. For nonvanishing  isospin chemical potential the pion system has only  $U(1)$ internal symmetry, therefore the field boundary condition  forms as well a circle $S^1$. Thus, the  map $S^1 \to S^1$ completely characterizes the behavior of the vortex at the boundary.  Since this map  cannot be smoothly deformed  to the trivial map, it follows that the vortex solution is topologically stable.}

\subsection{Stress-energy tensor}
The thermodynamic and mechanical properties of the system are determined by the
the stress-energy tensor 
\be
 T^\mu_\nu = \frac{\delta {\mathcal L} }{\delta\partial_\mu \alpha} \partial_\nu \alpha +\frac{\delta {\mathcal L} }{\delta\partial_\mu \Phi} \partial_\nu \Phi  - \delta^\mu_\nu {\mathcal L} \,,
\ee
where we have assumed that the $\Theta$ field is fixed. It allows us
to obtain the energy density and 
 pressure components,  respectively given by 
\begin{widetext}
\begin{align}
\epsilon &= 2K\left[(\de_r \alpha)^2+\frac{1}{r^2}(\de_\varphi \alpha)^2+\sin^2\alpha\left((\de_r \Phi)^2 + \frac{1}{r^2}(\de_\varphi \Phi)^2\right)+\mu_I^2 \sin^2\alpha-2m_\pi^2(\cos\alpha -1)\right]\,, \label{eq:epsilon}\\
    P_{r} &= 2K \left[(\de_r \alpha)^2 -\frac{1}{r^2}(\de_\varphi \alpha)^2+ \sin^2\alpha\left((\de_r \Phi)^2-\frac{1}{r^2}(\de_\varphi \Phi)^2\right) +\mu_I^2 \sin^2\alpha + 2 m_\pi^2(\cos\alpha-1)\right]\,, \label{eq:Prr} \\
    P_{\varphi} &=2K \left[-(\de_r \alpha)^2+\frac{1}{r^2}(\de_\varphi \alpha)^2 - \sin^2\alpha\left((\de_r \Phi)^2-\frac{1}{r^2}(\de_\varphi \Phi)^2\right) +\mu_I^2 \sin^2\alpha + 2 m_\pi^2(\cos\alpha-1)\right]\,,\label{eq:Pff}\\
 P_{z} &= 2K \left[ -(\de_r \alpha)^2-\frac{1}{r^2}(\de_\varphi \alpha)^2 - \sin^2\alpha\left((\de_r \Phi)^2+\frac{1}{r^2}(\de_\varphi \Phi)^2\right) +\mu_I^2 \sin^2\alpha + 2 m_\pi^2(\cos\alpha-1) \right] \,,  
    \label{eq:Pzz}
\end{align}
\end{widetext}
where $\alpha$ and $\Phi$ are  assumed to be functions of $r$ and $\varphi$. Then, the isospin density 
\be\label{eq:nI}
n_I = \frac{\de P_{r}}{\de \mu_I}= \frac{\de P_{\varphi}}{\de \mu_I} = \frac{\de P_{z}}{\de \mu_I}= 4 K \mu_I \sin^2\alpha\,,
\ee
depends only on $\alpha$ and it is in general space dependent. 
The stress-energy tensor has as well off-diagonal components, in particular 
\begin{align}\label{eq:T_0phi}
T_{0 \varphi} & = \frac{4 K}{r}   \mu_I \sin^2\! \alpha\,  \de_\varphi \Phi  \,,\\
T_{0 r} & = 4 K  \mu_I \sin^2\! \alpha\,  \de_r \Phi\,,\label{eq:T_0r}
\end{align}
are the  momentum densities in the $xy$-plane.
The angular momentum with respect to the cylinder  $z$-axis is  given by 
\be
L_z= \int d^3 r (\bm r \times \bm T)_z\,, 
\ee
where $\bm T = (T_{0r}, T_{0\varphi})$. For completeness,
\be
T_{\varphi r}  =\frac{4 K}{r} \left[  \de_\varphi \alpha \de_r \alpha + \sin^2 \alpha \de_\varphi \Phi \de_r \Phi  \right]\,,\label{eq:T_phir}
\ee
is the shear stress, which should vanish  because the system is not viscous: we are assuming negligible temperature.

\section{Single vortex}
\label{sec:single}
We begin with the case of a single vortex imprinted to a non-rotating pion system.  If the vortex is at the center of the system and it is parallel to the $z$--axis it does not break the  cylindrical symmetry. In this case,  the potential can only depend on the radial coordinate. Since the potential is   a functional of  $\alpha$,   it follows that we have to assume that 
\be
\alpha \equiv \alpha (r)\,,
\ee
and thus the  BPS Eqs.\, \eqref{eq:SFBPS1}  and \eqref{eq:SFBPS2}  read
\begin{align}
\de_{r}\alpha \pm  \sin\alpha\, \frac{1}{r}\de_\varphi \Phi &=0\ ,
\label{CBPS1} \\
\sin\alpha\,\de_{r}\Phi &=0\,,
\label{CBPS2}
\end{align}
where the latter equation implies that  $\Phi \equiv\Phi(\varphi)$. 
Given the periodicity condition in Eq.\,
\eqref{eq:multivalued_Phi}, as we already discussed, there are two different class of solutions: $\Phi$ is a sinusoidal function of $\varphi$ or it is linearly dependent on it. The sinusoidal angular dependence is incompatible with  Eq.\,\eqref{CBPS1}, because $\alpha$ does not depend on $\varphi$. For this reason we can only consider solutions of Eq.\,  \eqref{CBPS2} of the form
\begin{equation}   \label{eq:Phin}
\Phi=n\,\varphi \,,
\end{equation}
where $n \in \mathbb{Z}$ is the winding number. In this way we have that Eq.\,\eqref{CBPS1} turns into
\begin{equation}\label{eq:bpsalpha}
    \frac{d \alpha}{dr} + \frac{n}{r}\sin \alpha =0\,,
\end{equation}
where we restrict to  consider the positive sign because $n$ can be a positive or negative integer.   The  above equation  is  invariant under the scaling $r \to \lambda\, r$, with $\lambda$ a constant, it is indeed independent of all the energy scales $m_\pi$, $\mu_I$ and  $f_\pi$. It has no information on interactions and masses because it  is   solely determined by the derivative term.  
\begin{figure}[t!]
\centering
    \includegraphics[width=\columnwidth]{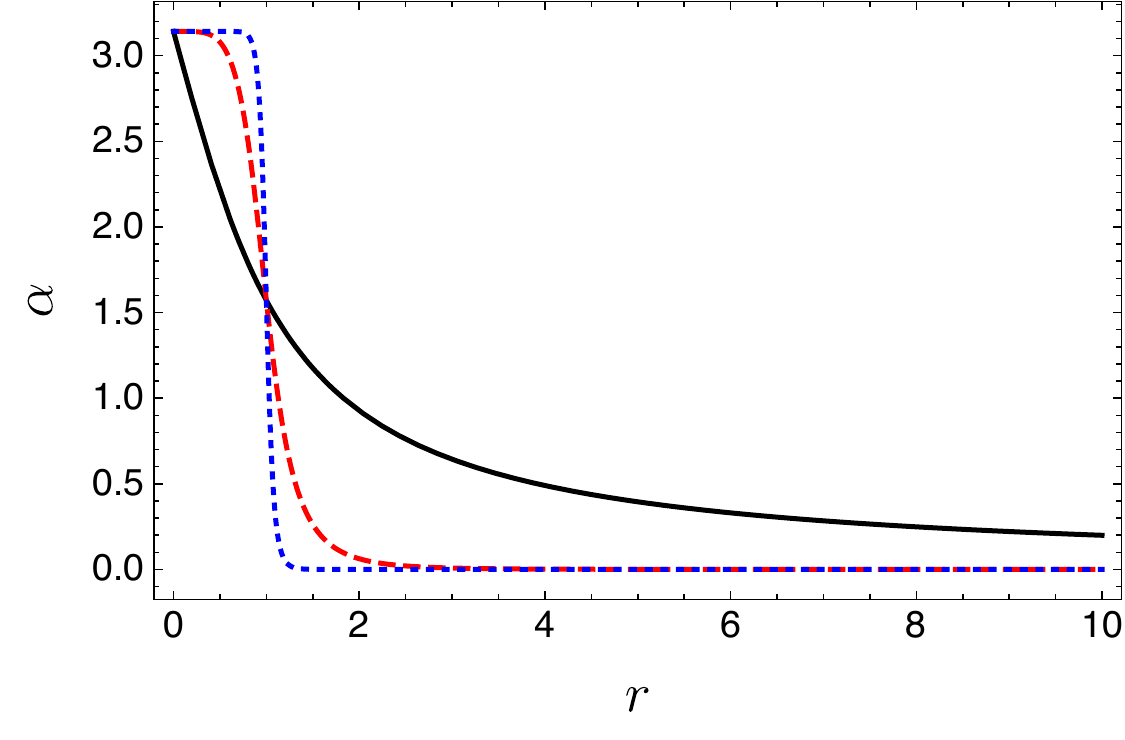}
    \caption{Modulation of the $\alpha$ function  determined by a single vortex at the center, see Eq.\,\eqref{eq:alpha_single}, as a function of the  distance from  the $z$-axis of the cylinder  (in units of  $1/m_\pi$).   The reported results are obtained with winding number $n=1$, solid black line, $n=5$, dashed red line and $n=20$, dotted blue line. All the lines intersect at $r=1, \alpha = \pi/4$.     }
    \label{fig:alpha}
\end{figure}

The  solution of the  differential equation \eqref{eq:bpsalpha} is 
\begin{equation} \label{eq:alpha_single}
    \alpha (r) = 2 \arctan (C^n r^{-n})\,,
\end{equation}
which depends on the integration constant, $C$, that  determines  the typical length scale variation of $\alpha$. In the following we will set $C=1/m_\pi$, and thus all length scales will be expressed in units of $1/m_\pi$. Although arbitrary, this is somehow a natural choice, given that 
in the  leading order $\chi$PT Lagrangian  $f_\pi$ factorizes and $\mu_I\sim m_\pi$.  
In Fig.\,\ref{fig:alpha},  we report the behavior obtained with $n=1$, $n=5$ and $n=20$. All the lines start from $\alpha= \pi$ for $r=0$,  intersect at $r=1$, $\alpha=\pi/2$, and  tend to $0$ for large $r$.  With increasing $n$, the variation of  $\alpha$ around  $r=1$ is more pronounced, indeed  it tends to the Heaviside step function $\Theta_H(1-r)$ for $n\to \infty$.  The region $r <1$ corresponds to momenta larger than $m_\pi$ meaning that the used leading order $\chi$PT Lagrangian is not adequate to describe it. This is also signaled by the fact that in the homogeneous phase we have seen that  $\alpha \in [0,\pi/2)$, in that case values of $\alpha$ larger than $\pi/2$ correspond to imaginary isospin chemical potentials.

Upon substituting Eqs.\,\eqref{eq:Phin} and \eqref{eq:bpsalpha} in Eqs.\,\eqref{eq:epsilon}, \eqref{eq:Prr}, \eqref{eq:Pff} and \eqref{eq:Pzz}, the energy density and pressure components are respectively given by
\begin{align}
\epsilon &= 2 K \left[\sin^2\!\alpha\left(\mu_I^2 + \frac{2 n^2}{r^2}\right) - 2 m_\pi^2 (\cos\!\alpha -1)\right] \,,\nonumber\\
P_{r}&=P_{\varphi}=  2 K \left[\mu_I^2\sin^2\!\alpha + 2 m_\pi^2(\cos\!\alpha -1) \right]\,,\nonumber\\
P_{z}&= 2K  \left[\left(\mu_I^2 - \frac{2 n^2}{r^2}\right)\sin^2\!\alpha + 2 m_\pi^2 (\cos\!\alpha -1) \right]\,,
\label{eq:central_thermo}
\end{align}
meaning that the pressure is isotropic in the $xy$-plane.

The isospin number density in Eq.\,\eqref{eq:nI} takes the  expression 
\be\label{eq:nIn}
n_I =4 K  \mu_I \sin^2\alpha = 16 K \mu_I \frac{r^{2 n}}{(1+r^{2 n})^2}\,,
\ee
which  is invariant under $n\to -n$; in other words, the vortex and the antivortex solutions are both independent of $\varphi$ and have the same radial dependence. The isospin number density vanishes at large distances as $r^{2n}$. This behavior is determined by the fact that $\alpha$  tends to $0$ for large values of $r$. Notice that  it is not possible to consider a vortex as a perturbation  of a homogeneous background: the number density is completely determined by the vortex. 
A homogeneous background corresponds to $\alpha$ constant,  thus to consider a vortex on  top of it one should replace $\alpha \to \alpha + \text{A}$, where $A$ is some  constant. However, the only solution compatible with  the BPS Eq.\,\eqref{eq:bpsalpha} is the one with $A=0$.

Regarding the off-diagonal elements of the stress-energy tensor, the shear stress in Eq.\,\eqref{eq:T_phir} vanishes, therefore the system is inviscid. The component $T_{0r}$ in Eq.\,\eqref{eq:T_0r} vanishes as well, meaning that there is no radial flow, while  from  Eq.\,\eqref{eq:T_0phi} we have that
\be\label{eq:T0phi}
T_{0 \varphi} = n\frac{4 K}{r} \mu_I \sin^2\alpha = n \frac{n_I}{r}\,,
\ee
which corresponds to the momentum density along the tangential direction.  These results imply that  the pion gas is  spinning with tangential velocity $v_\varphi \propto 1/r$  as appropriate for a quantized vortex in a superfluid. From Eq.\,\eqref{eq:T0phi} we can calculate  the angular momentum along the $z$-axis, which  is given by 
\be\label{eq:Lzn}
L_z = \int d^3 x T_{0 \varphi} r = n N_I\,,
\ee
where
\be\label{eq:NI}
 N_I =\int d^3x n_I\,,
\ee
is the total isospin number. 
In other words, the angular momentum per particle is  $\hbar$ as in standard superfluids.

\subsection{Global BPS condition}

The components of the stress-energy tensor  depend on $\mu_I$, which cannot take arbitrary values. The reason is that we are asking that  the BPS condition in Eq.\,\eqref{eq:BPS_condition} be satisfied.  Considering a vortex  at the center of the cylinder and assuming a constant  isospin chemical potential,  we obtain that 
\be\label{eq:BPS_R}
\mu_I^2  = 2 m_\pi^2 \frac{\int_0^R dr r (1-\cos\alpha)}{\int_0^R dr r \sin^2{\alpha}}\,,\ee
which implies that  $\mu_I$  depends on the behavior of the condensate across the whole system.
We shall refer to  equations like \eqref{eq:BPS_R} as   global BPS conditions. It is important to remark that this means that the system is not at a stationary point of the action, therefore Eqs.\,\eqref{eq:Phin} and \eqref{eq:bpsalpha} are 
 not in general a solution of the 
Euler-Lagrange equations obtained from the Lagrangian in Eq.\,\eqref{eq:L_sempl}. 
Although corresponding to an unstable configuration, we find instructive this case: as we will see, it has some interesting features.

\begin{figure}[t!]
   \centering
    \includegraphics[width=\columnwidth]{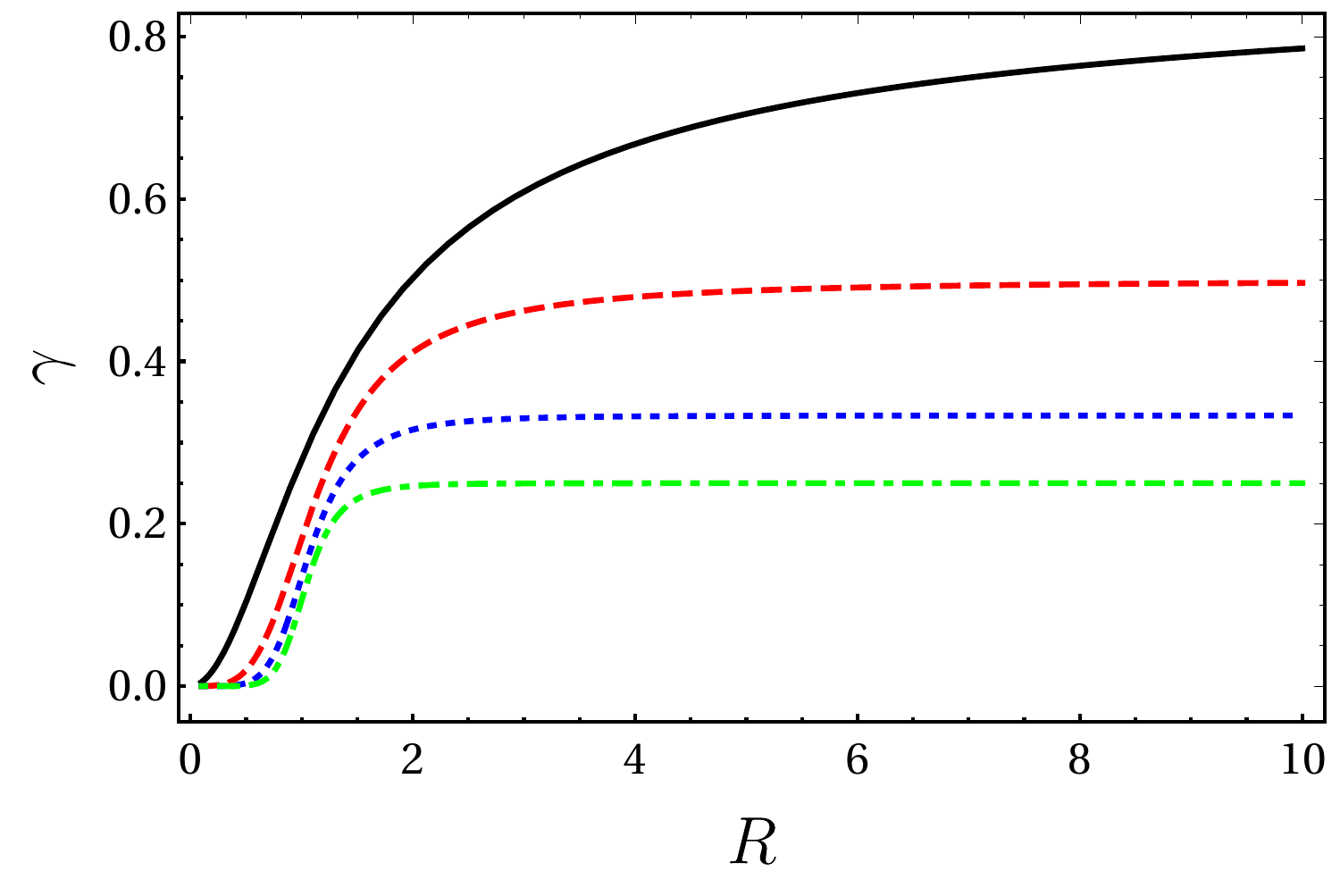}
            \caption{Values of  $\gamma=m_\pi^2/\mu_I^2$ that satisfy the global BPS condition, see Eq.\,\eqref{eq:BPS_R},           as a function of the  distance from  the $z$-axis of the cylinder  (in units of  $1/m_\pi$). The reported results  are obtained with  winding number $n=1$, solid black line, $n=2$, dashed red line, $n=3$, dotted blue line, and $n=4$, dot-dashed green line.}
    \label{fig:gamma_R}
\end{figure}

Let us  consider a central vortex with minimal nontrivial winding $n=1$. Using  Eq.\,\eqref{eq:alpha_single} we have that 
\be
\int_0^R dr r \sin^2{\alpha} = 2 \left[ -\frac{R^2}{1+R^2} + \log(1+R^2)\right]\,,
\ee
while 
\be
\int_0^R dr r (1-\cos\alpha)= \log(1+R^2)\,,
\ee
therefore we obtain the analytical expression
\be\label{eq:gamma_R}
\gamma(R) = \frac{m_\pi^2}{\mu_I^2}= 1-\frac{R^2}{1+R^2} \frac{1}{\log(1+R^2)}\,,
\ee
  corresponding to the solid black line in  Fig.\,\ref{fig:gamma_R}. This function logarithmically converges to $1$ from below. 
This means that for any dimension of the system, it exists a value  $\mu_I > m_\pi$ that allows one to satisfy the global BPS condition. In the homogeneous phase we found that values of 
$\mu_I$ larger than  $m_\pi$ indicate  pion condensation,  with $\alpha$ given in Eq.\,\eqref{eq:alpha_hom}. Here, there is a modulation of $\alpha$, however in the  thermodynamic limit $R \to \infty$, such modulation, for $r>1$,  smoothens  and at the same time   we have that $\mu_I \to m_\pi$.  This suggests that for $\mu_I \gtrsim m_\pi$, the global BPS condition approximates the actual stationary point of the action.
The above reasoning can be extended to arbitrary winding number. In particular,  a vortex at the center of the system with arbitrary winding number, $n$,  satisfies  the global BPS condition if $\gamma\leq 1/n$, as shown in Fig.\,\ref{fig:gamma_R}.
 In other words, the vortex solution with winding number $n$  satisfying the global BPS condition can only appear for  $\mu_I \geq \sqrt{n} m_\pi$.

\begin{figure}[t!]
   \centering
   \includegraphics[width=\columnwidth]{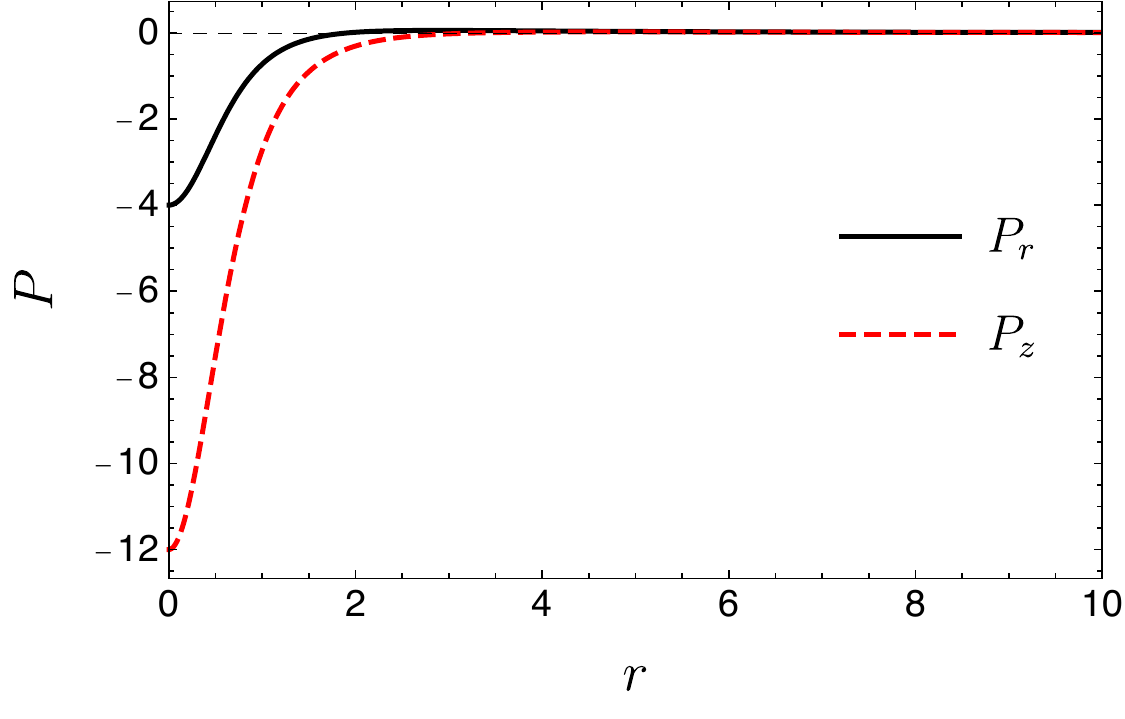}\\ \includegraphics[width=\columnwidth]{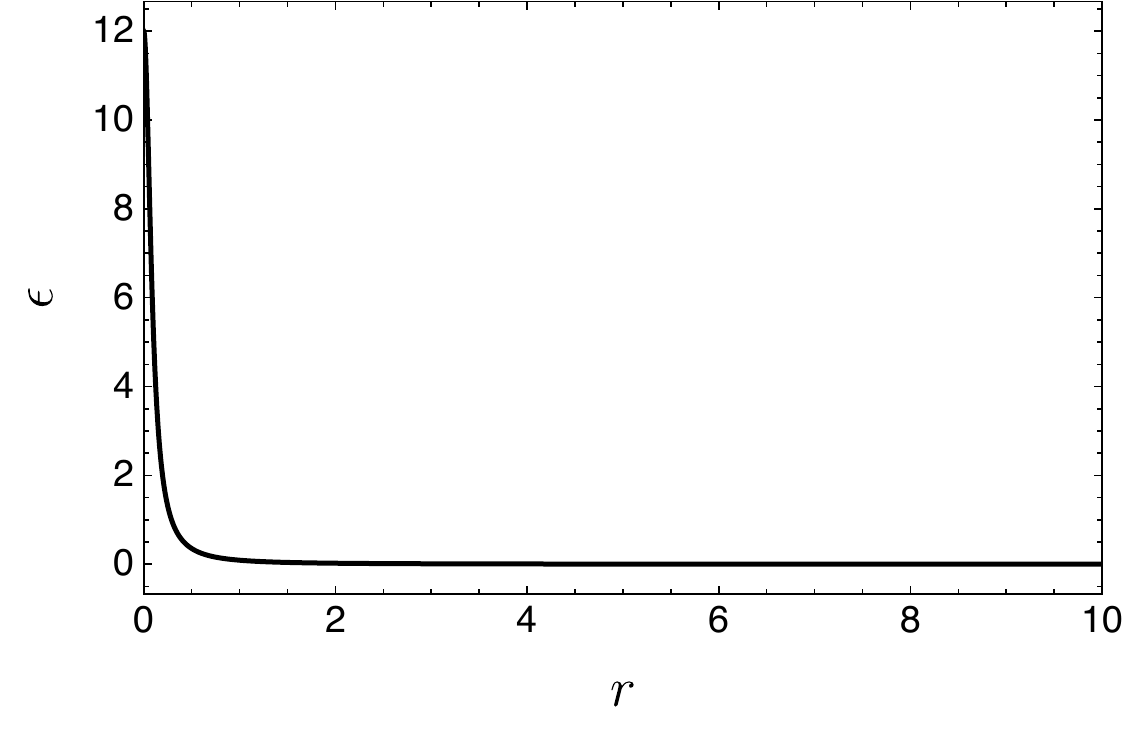}
    \caption{Pressure and energy density determined by a central vortex with winding number $n=1$ (in units of $2 K m_\pi^2$),       as a function of the  distance from  the $z$-axis of the cylinder  (in units of  $1/m_\pi$). 
        The results are  obtained using the global BPS condition, see Eqs.\,\eqref{eq:BPS_R} and \eqref{eq:gamma_R} with $R=10$,. 
            Top panel the pressure in the $xy$-plane is in black, while the pressure along the $z$--direction is the dashed red line. Bottom panel, energy density.  }
    \label{fig:pressure}
\end{figure}

Once the isospin chemical potential is known, we can compute the various thermodynamic quantities. 
We report in Fig.\,\ref{fig:pressure} pressure (top panel) and energy density (bottom panel) for  the case $n=1 $ and $R=10$. Results obtained with different values of $R$ are similar. Both  pressure components are negative at small distance from the cylinder axis, while they become positive at sufficiently large distances.  The pressure along   $z$ vanishes  at  $r \simeq 3.3$, while the pressure in the $xy$-plane vanishes  at  $r \simeq 1.9$. 
These values slightly increase with the system size $R$. Negative values of the pressure are typical of vortices: high fluid velocity produces a local depression. The negative pressure in $xy$--plane indicates that the vortex is willing to capture matter, the system is therefore  unstable unless some external force will prevent matter from falling in the vortex. Such behavior is a consequence of the fact that the considered solutions of the BPS equations \eqref{eq:BPS_R} do not correspond to an action stationary state.    
Negative values of the pressure in the $z$--direction are instead due to the vortex swirling: as in standard vortices the pressure along the vortex axis becomes large and negative close the core.

\begin{figure}[t!]
   \centering
    \includegraphics[width=\columnwidth]{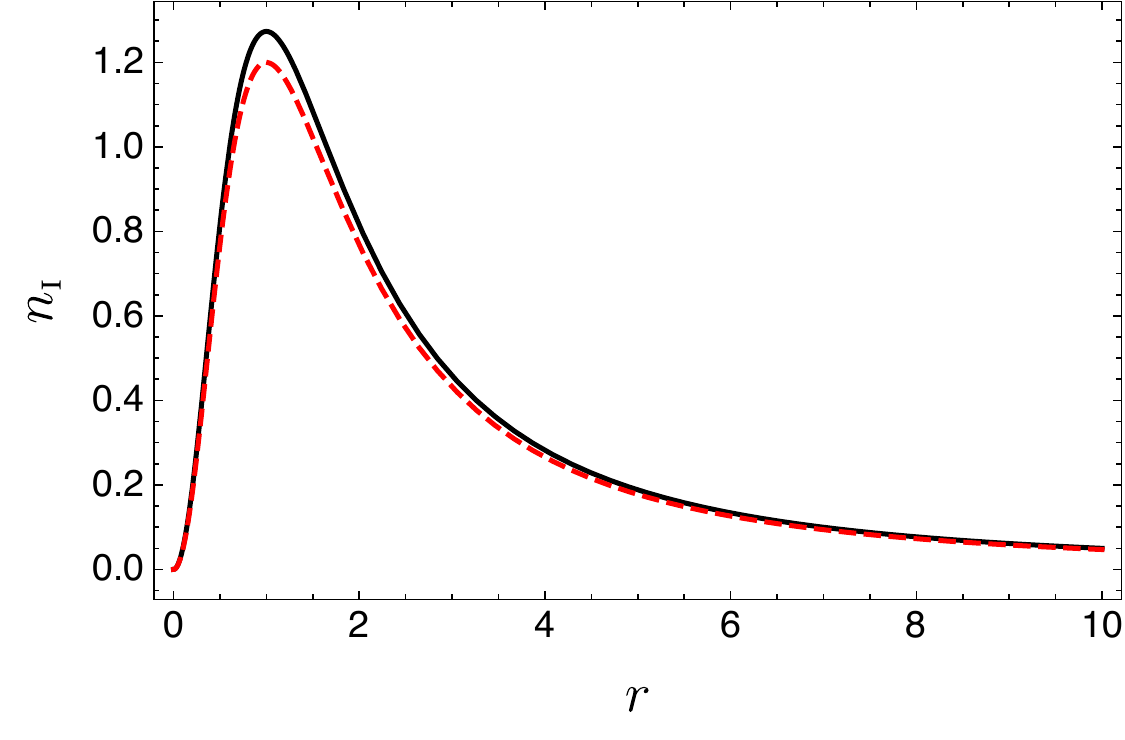}\qquad
            \caption{Radial dependence of the isospin number density 
                        for vortices with winding number  $n=1$,   $R=10$ (solid black)  and $R=20$ (dashed red) normalized to $16 K \mu_I$. Distances are in units of  $1/m_\pi$.}
    \label{fig:n_I}
\end{figure}

Negative values of the pressure close to the vortex core suggest an interpretation in terms of  a space dependent bag. As in the MIT bag model, the negative pressure may  serve to confine matter: pions cannot spread from the vortex.  For each value of $r$, the pressure  is constant that is independent of $\varphi$ and $z$.   This means that there is a sort of cylindrical confinement. In the MIT bag model the bag pressure is about $(140-150 \text{ MeV})^4$, here we find that it is of the order of $16 K m_\pi^2 = 4 f_\pi^2 m_\pi^2 \simeq m_\pi^4$, and thus of a similar magnitude. As in the MIT bag model, this negative  pressure comes with  a positive energy density, see the bottom panel in Fig.\,\ref{fig:pressure}. 
Moreover, both  pressure and energy density tend to  vanish for large $r$, meaning that confinement becomes negligible far from the place where  the number density is large. The number density is indeed peaked at $r=1/\sqrt{2}$, as can be seen from Fig.\,\ref{fig:n_I}, where we report $n_I$ for two different values of the radial size.   Such behavior of the number density is somehow similar to that of trapped ultracold atoms: the vortex core does indeed correspond to the region where the condensate vanishes, while the the trapping potential induces a vanishing number density  at the edge of the system. Here we observe a similar behavior, but the decrease of the number density far from the vortex core  is not due to an external potential: it is a consequence of  pion self-interactions. The expression of the number density comes from  the term proportional to $\sin^2 \alpha $ in the Lagrangian. At sufficiently large distances from the vortex core $\alpha \ll 1 $, see Fig.\,\ref{fig:alpha},  and  we can  Taylor expand the sine function. Taking into account   Eq.\,\eqref{eq:alpha_phi} we can  interpret the sinusoidal terms in $\alpha$  as a series of pion self-interactions which become more and more relevant as one approaches a vortex from large $r$. As already noted, this  calls for adding more terms to the $\chi$PT Lagrangian to describe the short distance behavior close to the vortex core.   
Combining Eq.\,\eqref{eq:gamma_R} with Eqs.\,\eqref{eq:nI} and  
\eqref{eq:NI} we find that the total isospin number is given by
\be\label{eq:NIglobal}
N_I = \frac{16 K \pi   \ell_z}{m_\pi^2 \sqrt{\gamma(R)}}\left(\log[1+R^2] - \frac{R^2}{1+R^2}\right)\,,
\ee
where  $\ell_z$ is the transverse dimension of the system measured in units of $1/m_\pi$. Since Eq.\,\eqref{eq:Lzn} holds, this is as well the total angular momentum.

\subsection{Local BPS condition}
\label{sec:L0}
In the homogeneous phase,  $\alpha$  and $\mu_I$ are linked by Eq.\,\eqref{eq:alpha_hom}. In a system in which  $\mu_I$ is a smooth function of $r$, one expects that  $\alpha$ is modulated as well. Conversely,  a space modulation of $\alpha$ should go along with a space modulation of $\mu_I$. Considering a space dependent isospin chemical potential allows us as well to overcome the main limitation of the global BPS condition:  Eq.\,\eqref{eq:BPS_R} ensures that the free-energy contribution of the potential term vanishes, but it does not correspond to a stationary point of the action. We can  define a local BPS condition such that the free energy vanishes at every point.   This happens if the isospin chemical potential is space dependent and takes the expression 
\be\label{eq:mu(r)}
\mu_I^2 = 2 m_\pi^2 \frac{1-\cos\alpha}{\sin^2{\alpha}}\,,
\ee
such that $F_V = 0$ at every point. 
{In this case 
 the solutions of the BPS equations 
are as well solutions of  the Euler-Lagrange equations,
\begin{align}\label{eq:EL1}
\sin^2 \alpha \de^\mu \de_\mu \Phi + \sin (2 \alpha) \de_\mu \alpha \de^\mu \Phi &=0\,,\\
\label{eq:EL2}
\de_\mu\de^\mu \alpha - \sin\alpha \cos\alpha \partial_{\mu}\Phi\partial^{\mu}\Phi & =0\,,
\end{align}
the system is indeed at   an action stationary point. Although general, this result can be easily shown  for a single vortex at the center. 
Regarding Eq.\,\eqref{eq:EL1}, since $\alpha \equiv \alpha(r)$ and $\Phi \equiv \Phi(\varphi)$ it follows that $ \de_\mu \alpha \de^\mu \Phi =0$. Given Eq.\,\eqref{eq:Phin},  we also have  $\de^\mu \de_\mu \Phi =0$,  therefore
Eq.\,\eqref{eq:EL1} is satisfied. 
Regarding Eq.\,\eqref{eq:EL2}, substituting  Eq.\eqref{eq:Phin}, we have that
\be\label{eq:EL2_n}
\nabla^2\alpha = \frac{n^2}{r^2}\sin\alpha \cos\alpha \,,
\ee
which can be proven to be satisfied by the BPS solution by  repeatedly differentiating  Eq.\,\eqref{eq:bpsalpha}; in this way we obtain that
\be\nabla^2 \alpha= -n\frac{1}{r}\de_r \sin\alpha = -n\frac{1}{r}\cos\alpha \de_r \alpha = \frac{n^2}{r^2}\sin\alpha \cos\alpha \,,
\ee
which indeed agrees  with \eqref{eq:EL2_n}. Therefore, the BPS solution in Eqs.\,\eqref{eq:Phin} and \eqref{eq:alpha_single} are solutions of the Euler-Lagrange equations.
}

Taking into account Eq.\,\eqref{eq:alpha_noncentral} we have that

\be\label{eq:muI_r}
\mu_I^2 = m_\pi^2\left(1 + \frac{1}{r^{2 n}}\right)\,,
\ee
therefore it diverges at the vortex center. However, as already notice, the used $\chi$PT leading order Lagrangian is valid for chemical potentials not much larger than the pion mass. For this reason  we assume that the above solution is valid only for  $r \geq 1 $,
corresponding to $\mu_I \leq \sqrt{2}\, m_\pi $. The inclusion of higher order terms in the $\chi$PT Lagrangian should regularize  the vortex core. 
\begin{figure}[t!]
    \centering
    \includegraphics[width=\columnwidth]{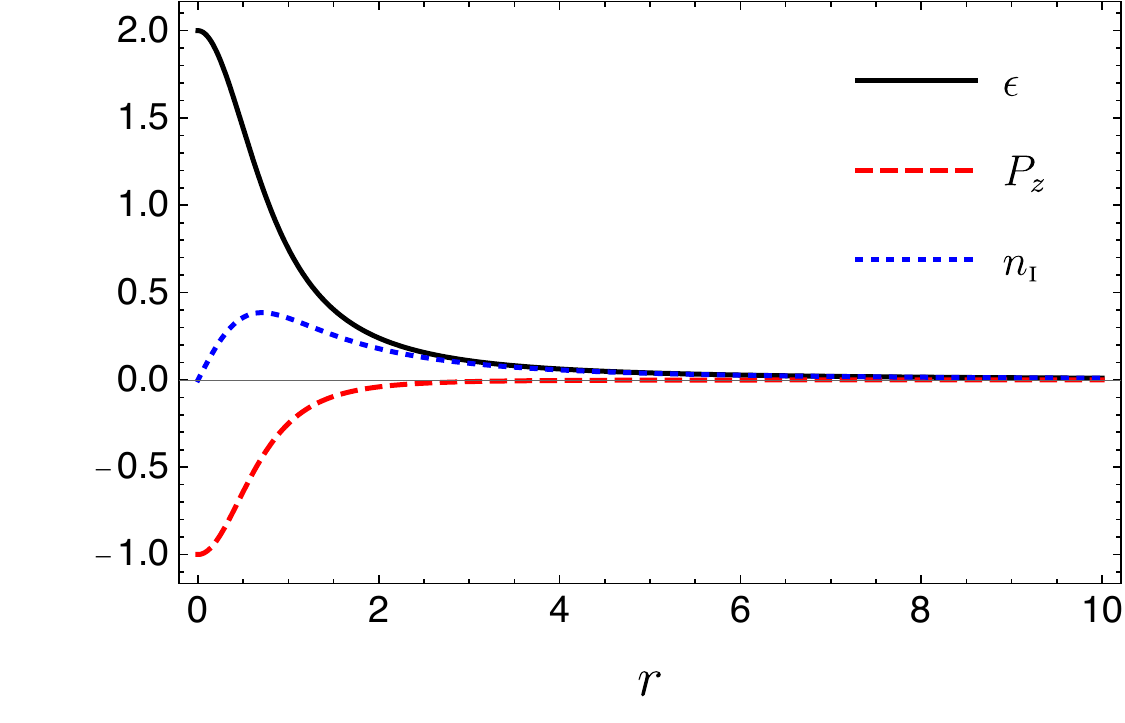}
    \caption{Radial dependence of the energy density, pressure (in units of $16 K m_\pi^2$) and number density (in units of $16 K m_\pi$). Results  obtained  with a  central vortex with winding number $n=1$ using the local BPS condition, see the main text for more details. Distances are in  units of  $1/m_\pi$.}
    \label{fig:ePn}
\end{figure}

From Eq.\,\eqref{eq:mu(r)}, we have that
\be
\alpha = 2 \arctan\left(\sqrt{\frac{\mu_I^2}{m_\pi^2}-1}\right)\,,
\ee
which can be seen as a combination of  the local BPS condition with the BPS solution in Eq.\,\eqref{eq:alpha_single}. Upon substituting these expressions in Eq.\,\eqref{eq:central_thermo}, we obtain that
\begin{align}
    \epsilon&= 16Km_{\pi}^2\frac{2+r^2}{(1+r^2)^2}\,,\nonumber\\
    P_{r}&=P_{\varphi} =0\,,\nonumber\\
P_{z} & = -16Km_{\pi}^2\frac{1}{(1+r^2)^2}\,,\nonumber\\
n_I & = 16Km_{\pi}\frac{r}{(1+r^2)^{3/2}}\,,
\end{align}
therefore the pressure in the $xy$-plane is isotropic, homogeneous and equates the vacuum pressure.  In Fig.\,\ref{fig:ePn} we show the radial dependence of $\epsilon$ (black solid line), $n_I$ (dotted blue line) and $P_z$ (dashed red line). Note that  $n_I$ is peaked at short distance from the vortex core, therefore the sort of confinement effect found using the global BPS condition still holds. Since the pressure in the $xy$-plane vanishes, the vortex does not tend to accrete  pions from the surrounding medium and it is therefore stable.
Although we are considering a stationary point of the action,   $P_z$ is negative. As already observed, this is a typical effect due to the vortex swirling.  In agreement with the results obtained using the global BPS case, we checked that the viscous component vanishes, that is $T_{r\varphi}=0$. 

\begin{figure}[t!]
    \centering
    \includegraphics[width=\columnwidth]{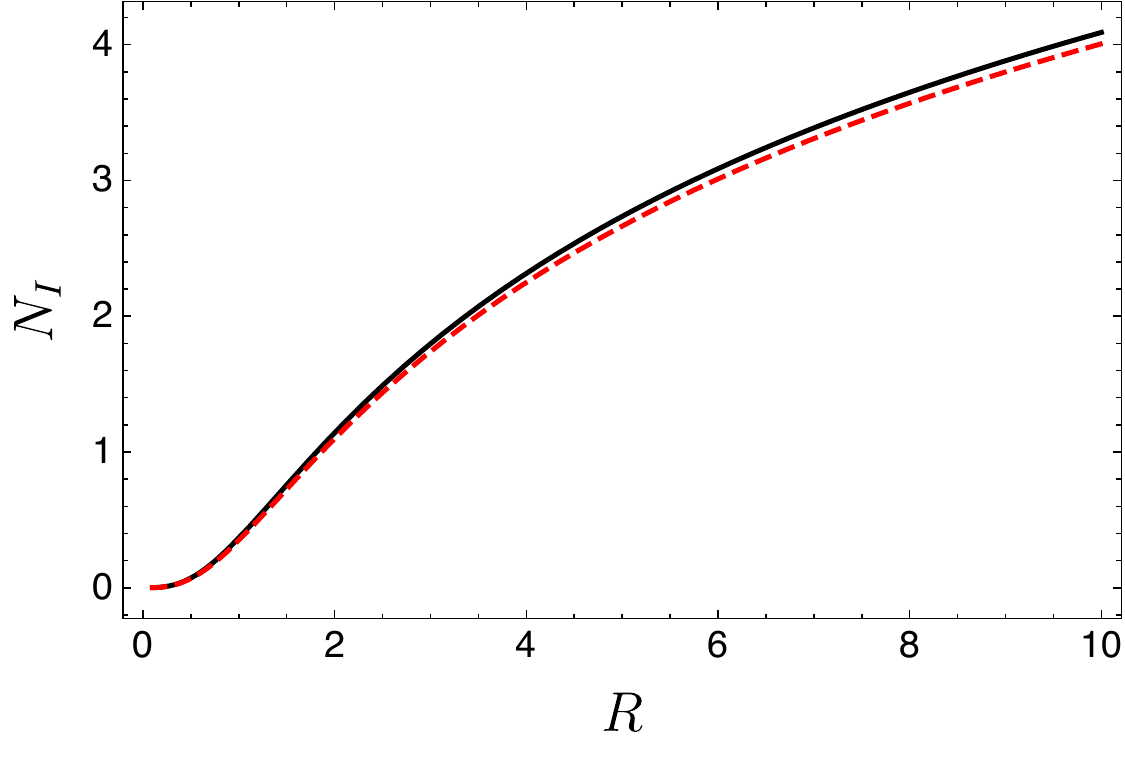}
    \caption{Total isospin number in presence of a  central vortex as a function of the cylinder radius, $R$. The isospin number  is in units of  $16\pi \ell_z K / {m_\pi^2} $, while the radius is in units of $1/m_\pi$. The solid black line is obtained with the global BPS condition, while the dashed red line using the local BPS condition.   }
    \label{fig:NItot}
\end{figure}

The angular momentum takes the same expression reported in Eq.\,\eqref{eq:Lzn}, that is $L_z = n N_I$, but now the total number of particles is given by
\be\label{eq:NIlocal}
N_I= \int d^3 x n_I =   \frac{32\pi K \ell_z}{m_\pi^2}    \left( \operatorname{arcsinh}(R) -\frac{R}{\sqrt{1+R^2}}\right)\,,
\ee
that is a positive and monotone function of $R$. In Fig.\,\ref{fig:NItot}, we compare the total isospin number obtained with the  global (solid black) and local (dashed red) BPS conditions. The two lines are  very similar  because in both cases the dominant contribution  comes from the derivative  part of the Lagrangian.

\section{More general vortex configurations}
\label{sec:general_vortex}
The previously obtained results can be extended to more general vortex configurations. We first consider  a noncentral vortex and then the most general case of an arbitrary number of vortices.

 \subsection{Noncentral vortex}\label{sec:noncentral}
When the vortex is not at the center of the system the cylindrical symmetry is broken. However, one can generalize the central vortex solution by taking
\be\label{eq:alpha_noncentral}
\alpha = 2 \arctan\left( |\bm r - \bm r_\text{v}|^{-n}\right)\,,
\ee
and
\be
\Phi =n\arctan \left( \frac{y-y_\text{v}}{x-x_\text{v}}\right) \,,
\label{eq:Phi_noncentral}
\ee
where $\bm r_\text{v} = (x_\text{v},y_\text{v})=r_\text{v}(\cos{\varphi_\text{v}},\sin{\varphi_\text{v}})$ is the vortex position and $n$ its winding number. Upon substituting these expressions in Eq. \eqref{eq:BPS_condition},  the global BPS condition is satisfied when
\be
    \gamma (R,r_\text{v})=\frac{-R^2-\chi+\chi \log \left(\frac{r_\text{v}^2 \left(R^2-r_\text{v}^2+1+\chi\right)}{-R^2+r_\text{v}^2-1+\chi}\right)+r_\text{v}^2+1}{\chi \log \left(\frac{r_\text{v}^2 \left(R^2-r_\text{v}^2+1+\chi\right)}{-R^2+r_\text{v}^2-1+\chi}\right)},
\ee
where 
\begin{eqnarray}
    \chi(R,r_\text{v})=\sqrt{R^4-2 R^2 \left(r_\text{v}^2-1\right)+\left(r_\text{v}^2+1\right)^2}\,.
\end{eqnarray}
This result is independent of the polar position of the vortex,  as expected.   In the limit $r_v \rightarrow 0$,
the central vortex result in Eq.\,\eqref{eq:gamma_R}  is recovered. Regarding the local BPS condition, the space dependent isospin chemical potential is obtained substituting Eq.\,\eqref{eq:alpha_noncentral} in Eq.\,\eqref{eq:mu(r)}. This ensures that $F_V=0$ at every point. 

\begin{figure}[t!]
    \centering
    \includegraphics[width=\columnwidth]{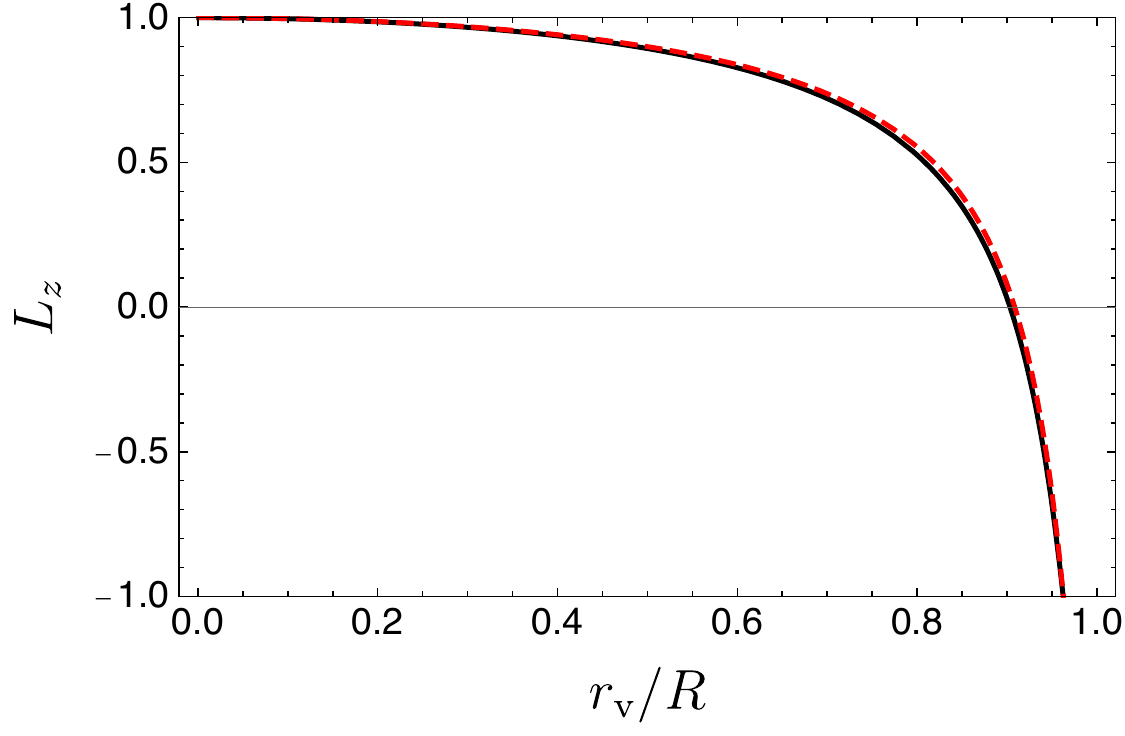}
    \caption{Angular momentum 
    generated by a single vortex  at a distance $r_\text{v}$ from the $z$-axis of the cylinder.
    The solid black line refers to the results obtained with the  global BPS condition, while the  dashed red line is obtained using the local BPS condition.}
    \label{fig:Ltot_g_l}
\end{figure}

The two BPS conditions give similar results as   is  evident in Fig.\,\ref{fig:Ltot_g_l}, where we show the dependence of the angular momentum on the distance of the vortex from the $z$--axis. The two lines correspond to the results obtained with the global (solid black) and local (dashed red) BPS conditions. Quite remarkable is the fact that the angular momentum changes sign if the vortex is very close to the boundary. This effect depends on the peculiar  isospin number density produced by a vortex, which is peaked in a region close to the vortex and vanishes both at the vortex core and far from it. When the vortex approaches the boundary, a large fraction of pions rotates clockwise giving a negative contribution to the angular momentum. The angular momentum reduction close to the boundary is  observed as well in systems of superfluid ultracold atoms. However, in that case  the angular momentum vanishes without changing sign when the vortex reaches the boundary.

 \subsection{Multivortex configurations}\label{sec:multi.vortex}
The single noncentral vortex solution in Eqs.\,\eqref{eq:alpha_noncentral} and  \eqref{eq:Phi_noncentral} can be extended to multivortex configurations.  To this end,   we rewrite the BPS equations \eqref{eq:SFBPS1} and \eqref{eq:SFBPS2}  as 
\be \de _{2}\Phi =\mp \frac{\de _{1}\alpha }{\sin \alpha },
 \quad
\de _{1}\Phi =\pm \frac{\de _{2}\alpha }{\sin \alpha },
\label{eq:multi}
\ee%
differentiating the first equation by $\de_1$,  the second by $\de_2$
and combing them one arrives at the following nonlinear differential equation %
\begin{equation}
\nabla^2 \alpha -\cot \alpha (\bm
\nabla \alpha )^{2}=0\,,  \label{solo1}
\end{equation}%
which   by means of  the  change of
variable%
\begin{equation}
\alpha =2\arctan %
\left( {\exp X}\right) \,,  \label{eq:solo2}
\end{equation}%
can be rewritten as 
\begin{equation}
\sin\alpha
\nabla^2 X = 0\,.  \label{solo3}
\end{equation}%
Since at the BPS point $\alpha$ does not identically vanish,    Eq. \eqref{solo1} is \textit{exactly
equivalent} to the Laplace's equation%
\begin{equation}
\nabla^2 X=0\,,
\label{eq:lapaceX}
\end{equation}%
while  \eqref{eq:multi} turns  into
\begin{eqnarray}
\ \de _{2}\Phi  =\mp \de _{1}X\,,  \quad
\ \de _{1}\Phi  &=&\pm \de _{2}X\,,  \label{eq:solo}
\end{eqnarray}%
meaning   that
$X$ and $\Phi $ are two conjugated harmonic functions.
At the center of a vortex, $\alpha = \pi$, see Fig.\,\ref{fig:alpha}, thus  from Eq.\,\eqref{eq:solo2} it follows that $X$ is singular.   In two dimensions, the most general harmonic function with singularities
is the Green function %
\be
X =-\sum_{j}^{N}n_{j}\log \left( |\bm r - \bm r_j| \right) \,,  \label{solo6} 
\ee%
where $N$ is the number of vortices, while  $\bm r_j = (x_j,y_j)$ and $n_j$  are respectively the  position in the $xy$-plane and winding number of the $j$-th vortex. From Eq.\,\eqref{eq:solo2} we then obtain
\begin{align}\label{eq:alpha_gen}
\alpha &= 2 \arctan\left(\prod_{j=1}^N |\bm r - \bm r_j|^{-n_j}\right)\,,
\end{align}
while from Eq.\,\eqref{eq:solo} we  find
\begin{align}
\Phi &=\sum_{j}^{N}n_{j}\arctan \left( \frac{y-y_{j}}{x-x_{j}}\right) \,,
\label{eq:Phi_gen}
\end{align}
which generalize the previously found  single vortex solution Eqs.\,\eqref{eq:alpha_noncentral} and  \eqref{eq:Phi_noncentral}.
These equations  have  several interesting features:
Eq.\,\eqref{eq:Phi_gen}  coincides
 with the Feynman-Onsager ansatz for the phase of superfluid multivortex configurations,
see\,\cite{Schmitt:2014eka, Pitaevskii_Stringari_2016} and references therein.
  Even if there are  many vortices with arbitrary winding number, thus making $X$  large in absolute value, $\alpha$  will always be in the $(0,\pi]$ interval.  Moreover, despite the highly nonlinear interactions present in $\chi$PT, the  BPS point  manifests a linear composition law. The
reason is that $X$ satisfies the  Laplace equation \eqref{eq:lapaceX}, thus  given an arbitrary number of solutions, their linear combination is still a solution. In particular, if 
\begin{align}
X_{1} &=\sum_{j}^{N_{1}}n_{j}\log \left( |\bm r-\bm r_j| \right)\,,  \\
X_{2} &=\sum_{k}^{N_{2}}n_{k}\log  \left( |\bm r-\bm r_k| \right) \,,
\end{align}%
are two solutions of the Laplace equation,  
then $\alpha_{1,2} = 2\arctan \left( {\exp X_{1,2}}\right)  $ are two solutions of Eq.\,\eqref{solo1} with $N_1$ and $N_2$ vortices, respectively.  Then, their nonlinear combination 
\begin{align}
\alpha_{1\oplus 2}= & 2\arctan \left( {\exp }\left( a {X}_{1}+b X_{2}\right)\right) \nonumber \\
 =& 2 \arctan\left(\prod_{j=1}^{N_1} |\bm r - \bm r_j|^{ -2 a n_j}\prod_{k=1}^{N_2} |\bm r - \bm r_k|^{-2 b n_k} \right) \nonumber \\
=& 2 \arctan\left(  \prod_{l=1}^{N} |\bm r - \bm r_j|^{- 2 n_l}\right)\,,
\end{align}
and
\be
\Phi_{1\oplus 2}= \sum_l^N n_l \arctan \left( \frac{y-y_{l}}{x-x_{l}}\right)\,,
\ee
are still valid solutions  with $N=N_1+N_2$ vortices having winding numbers $n_l = a\, n_j$ for $l\leq N_1$ and   $n_l = b\, n_k$ for $N_1 <l\leq N_2$.  Imposing the periodic boundary conditions in Eqs.\,\eqref{eq:multivalued_alpha} and \eqref{eq:multivalued_Phi}, only
integer values of $a$ and $b$ result admissible.

\section{Free energy and critical rotation frequency}
\label{sec:free-energy}
At the BPS point the system is not sensitive to the actual form of the potential, because the contribution of the potential free energy by definition vanishes.  
This means that the free energy can only depend on the geometry and topology of the system, in other words the system size and  winding number.

The free energy  for a configuration of one single vortex at the center has the analytical expression
\be\label{eq:F_R}
{\cal F} = n \frac{8 \pi K \ell_z}{m_\pi}\frac{R^2}{1+R^2}\,,
\ee
which is obtained combining Eqs.\,\eqref{eq:Fk_2}, \eqref{eq:Phin} and  \eqref{eq:alpha_single}. 
This  result differs significantly from the standard, approximate expression obtained in superfluids\,\cite{Stringari2016bec}. In that case the free energy 
diverges logarithmically with  $R$ and scales with $n^2$. Here the free energy is bounded:  at most it takes the value  ${\cal F}_\text{max}= n\,  8 \pi  K  \ell_z /m_\pi   $. This happens because the number density vanishes at large distances. More precisely, the free energy is dominated by the kinetic energy  of a vortex, which can be written as 
\be\label{eq:Ek}
E_k \sim \frac{1}{2} \ell_z \int_0^{2 \pi} d \varphi\int_0^R r dr  \rho v_s^2 \,, 
\ee
where $\rho$ is the mass density and 
\be\label{eq:vs} v_s \propto \frac{n}{r}\,, \ee is the superfluid velocity. In standard superfluids,  far from a vortex    $\rho$ tends to a constant value, thus the above integral yields
\be
 E_k \sim  n^2 \log(R) \,,
\ee
it is therefore  logarithmically divergent and scales with $n^2$.
In the present case the kinetic energy tends to a constant value at large $R$  because the isospin number density at large distances scales as $1/r^{2n}$, see Eq.\,\eqref{eq:nIn},  and then the  integral in Eq.\,\eqref{eq:Ek} becomes convergent.  The
linear scaling with  the winding number  is instead due to the fact that $\rho \sim n_I \sim r^{2n}/(1+ r^{2n})^2$,   defining $x=r^n$, one immediately finds that
\be\label{eq:change}
 \frac{d r}{r} = \frac{1}{n}\frac{d x}{x}\,,
\ee
and then substituting this expression in  Eq.\,\eqref{eq:Ek} and taking into account Eq.\,\eqref{eq:vs} it follows that  the energy cost of a vortex scales with $n$. This means that vortices with high winding number are not energetically unstable. Thinking of such configurations as made of $n$ overlapping vortices, each of them with winding number $1$, the free-energy cost  linearly scales  with the number of vortices, as if they were not interacting. Upon substituting Eqs.\,\eqref{eq:alpha_gen} and \eqref{eq:Phi_gen} in Eq.\,\eqref{eq:Fk} we have numerically verified  this aspect computing   the free energy of many vortices as a function of their relative distance. We find that configurations of $n$ vortices, all of them sufficiently close to the center, have the same free energy of a single vortex with winding number $n$. This  is  the analogous of the  behavior found in  superconducting vortices  at the critical point where type II superconductors turn into type I superconductors\,\cite{Bogomolny:1975de}.

\begin{figure}[t!]
    \centering
    \includegraphics[width=\columnwidth]{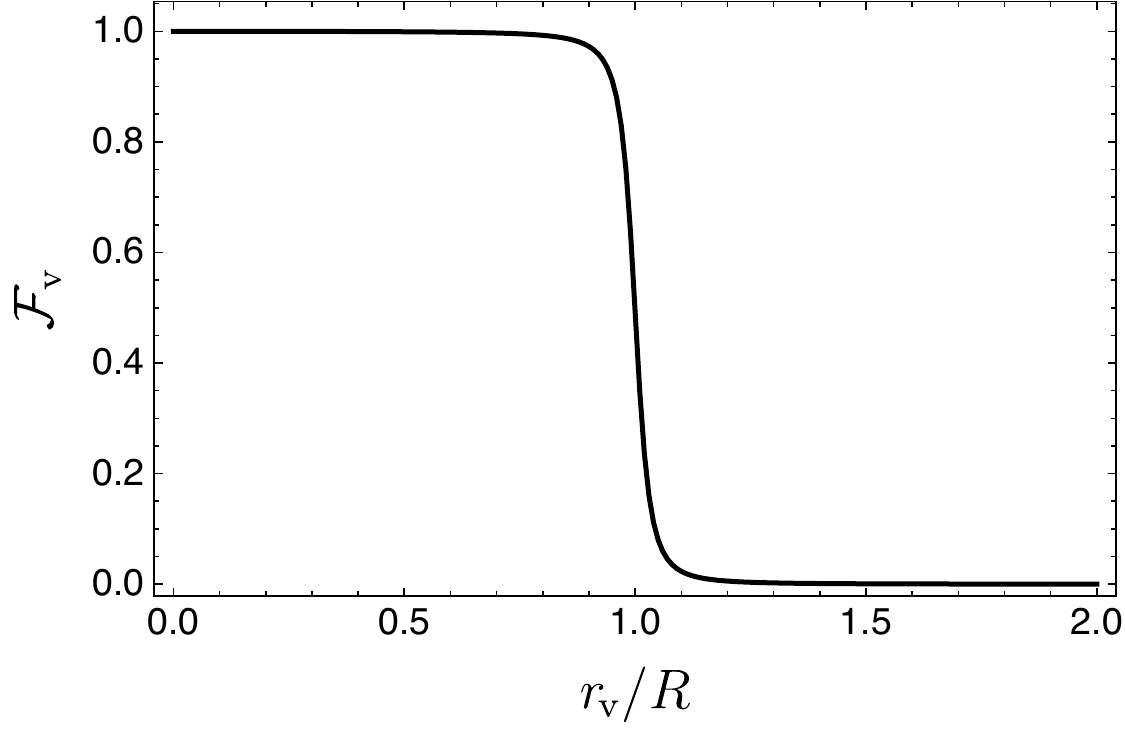}
    \caption{Dependence of the single vortex free energy as a function of its radial position. The free energy has been normalized to its value when the vortex is at the center, see Eq.\,\eqref{eq:F_R}.}
    \label{fig:F1}
\end{figure}

Given the symmetry of the system, the free energy of a vortex  depends only on its radial distance from the center.
We show  in Fig.\,\ \ref{fig:F1} the free energy of a vortex at a distance $r_\text{v}$ from the axis of the cylinder,    normalized to the value it takes when the vortex is at the center.   When the vortex is close to the center, its free energy is independent of its radial position. A vortex feels the boundary only when it is close to it, then  for $r_\text{v} = R$ its free energy becomes approximately one half of  the value in Eq.\,\eqref{eq:F_R}.  In summary, the free-energy cost of each vortex does not depend on the presence of  other vortices but only on its proximity to the boundary.

\begin{figure}[t!]
    \centering
    \includegraphics[width=\columnwidth]{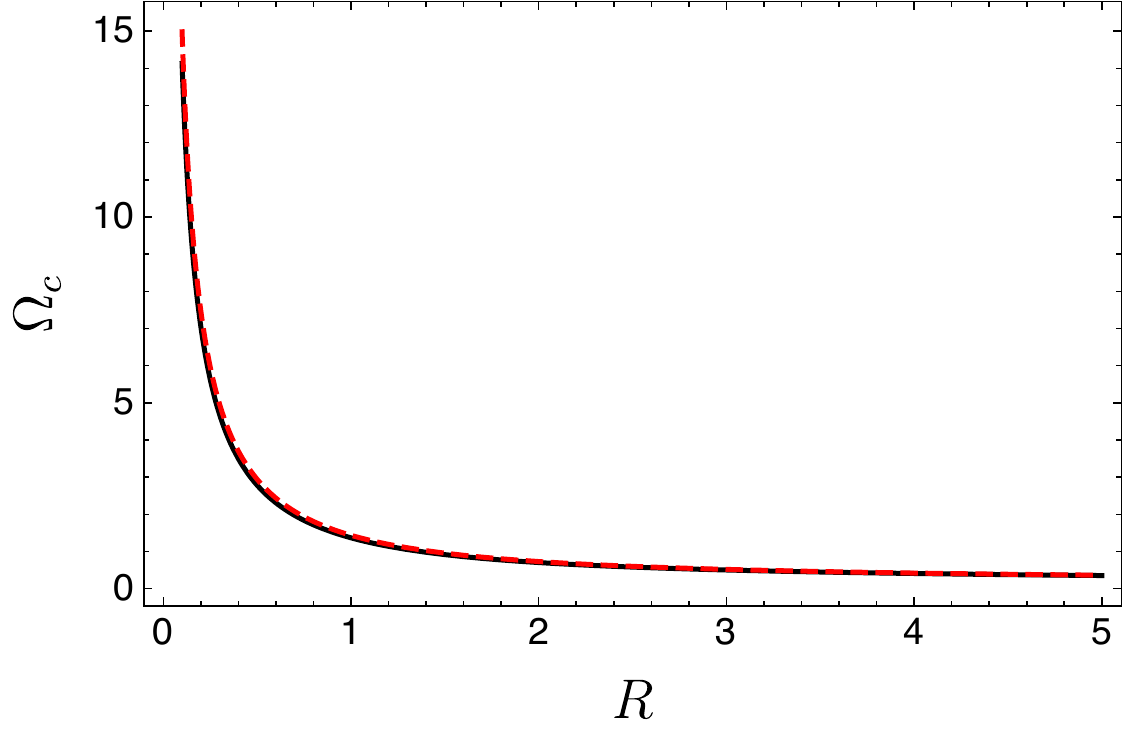}
    \caption{Estimated critical frequency, in units of $m_\pi$, for cylindrical  systems of different radii (in  units of  $1/m_\pi$). The black line is obtained using the global BPS condition, while the dashed red line with the local BPS condition.  }
    \label{fig:OmegaC}
\end{figure}

The  expression in Eq.\,\eqref{eq:F_R} allows us to estimate the critical frequency necessary for vortex nucleation. 
Suppose that the system is in a container rotating at an externally imposed rotation frequency $\Omega$.
 If the system is superfluid,  there should  exist a critical rotation frequency, $\Omega_C$, for vortex nucleation. We estimate it 
 by  the balance between the vortex energetic cost, as evaluated above,   and the kinetic energy gain occurring  when the superfluid  co-rotates with the container. This means that there is an additional free-energy contribution, not included in the above analysis.  
   Considering  the frame rotating at the externally imposed rotation frequency, the total free energy is 
\be\label{eq:Ftot}
{\cal F}_\text{tot} \simeq {\cal F} - \Omega L_z\,, 
\ee
where ${\cal F}$ is given in Eq.\,\eqref{eq:F_R}. This is an approximate expression because  we have evaluated the free-energy cost of a vortex in a nonrotating superfluid and then added the free-energy gain associated to rotation. In principle, one should evaluate ${\cal F}_\text{tot}$ starting from the $\chi$PT Lagrangian in a rotating frame; we will discuss this topic in a future publication. In the present paper we content ourselves with the approximate expression in Eq.\,\eqref{eq:Ftot}.  The condition of vortex nucleation at the trap  center is then  ${\cal F}_\text{tot}=0$, which gives
\be
\Omega_c = \frac{{\cal F}}{L_z}= \frac{{\cal F}}{N_I}\,,
\ee
where the total isospin number is in Eq.\,\eqref{eq:NIglobal} for the global BPS condition and in Eq.\,\eqref{eq:NIlocal} for the local BPS condition. In this approximation the critical frequency is proportional to  $m_\pi$ and scales with $R$ as shown in  Fig.\,\ref{fig:OmegaC}. 
The  critical frequency obtained with the global BPS condition almost matches that obtained with the local BPS condition because the free energy takes the same value in the two cases and the total isospin numbers  are very similar, see Fig.\,\ref{fig:NItot}. For a more realistic estimate of the critical frequency one should also take into account that vortices are typically generated at the boundary of the system. Given the radial dependence of the angular momentum reported in Fig.\,\ref{fig:Ltot_g_l}, dynamical vortex nucleation at the boundary is energetically disfavored because $L_z$ is negative,  meaning that the contribution $\Omega L_z$ in Eq.\,\eqref{eq:Ftot} is positive: it is a free-energy cost. Presumably, if a  vortex is dynamically nucleated at the boundary, it will rapidly move toward the center of the system: in this way $L_z$ becomes large and positive, hence  minimizing the total free energy.

\section{Conclusion}
\label{sec:conclusion}
We have analyzed the possible existence of quantized vortices in   $\chi$PT at nonvanishing isospin chemical potential.  Starting from the effective field Lagrangian in Eq.\,\eqref{eq:L_sempl} we have determined the minimal energy cost for  vortex nucleation at nonvanishing isospin chemical potential.    The obtained vortex configurations should then be understood as solutions of the low-energy effective field theory, combined with   the BPS critical condition. 
The used method is expected to be valid 
for isospin chemical potentials close to the pion mass. Actually, the BPS critical point is realized by appropriately tuning the isospin chemical potential and the system size. For definiteness we have assumed that the system has  cylindrical symmetry, eventually broken by noncentral vortices. This makes it similar to those realized in ultracold atom systems. However,  in this case, pions are self-confined by the vortex: the number density peaks in a region close to the vortex and tends to vanish both at the vortex core   and at the boundary. This suggests that the actual geometrical realization of the system is immaterial insofar the vortex is sufficiently far from  the boundary. 

Single vortex as well as multivortex states with quantized vorticity have  been  obtained. We have analyzed the single BPS vortex in detail. In particular, we have computed exactly the corresponding  energy density and momentum fluxes  as well as the quantized angular momentum. The existence of quantized vortices, as well as the vanishing of dissipative terms in the stress-energy tensor, unambiguously imply that the pion condensed phase is superfluid. In addition, we found that 
at the BPS point  the free energy is bounded and it linearly scales with the winding number. Therefore, vortices with large winding number are not unstable.  Such a result is expected to change when  perturbations to the BPS equations are included. This should allow us to check whether the obtained stationary solutions correspond to minima of the free energy. Work in this direction is underway.

The obtained results can in principle be tested by LQCD simulations in rotating frames\,\cite{Yamamoto:2013zwa}. We are not aware of any  such simulation  at nonvanishing $\mu_I$. By appropriately  tuning the isospin chemical potential,  vortices should appear close to the second order phase transition. Vortex nucleation at the boundary of these systems should happen as in standard superfluids,  then  followed by a  rapid  vortex drift toward the center of the cylinder.  If  doable, these simulations would be of  great interest, because they could indicate the possible generation of vortices of self-confined pions as well as the  existence of other interesting effects triggered by  rotation\,\cite{Huang:2017pqe, Zhang:2018ome, Evans:2025jsa}.

Whether our results could be relevant in astrophysics is an open question. Pion stars~\cite{Carignano:2016lxe,Brandt:2018bwq, Andersen:2018nzq, Stashko:2023gnn, Kirichenkov:2023omy, Cipriani:2024bcc, Chen:2024cxh} or neutron stars with  
  pion condensation in the core\,\cite{Mannarelli:2019hgn, Shapiro:1983du}    could  be two  types of  compact  objects  produced during the  collapse of  massive stars.  Such exotic compact stars are expected to rapidly cool  and  spin at high frequency,  hosting a large number of superfluid vortices. 
As we have shown, pion vortices  have large effect on the isospin number density, therefore they could  modify the stellar structure.

\section{Acknowledgments}
F.C. has been funded by Fondecyt grants No. 1240048, 1240043, 1240247 and by Grant ANID EXPLORACIÓN 13250014. The Centro de Estudios Cientificos (CECs) is funded by the Chilean Government through the Centers of Excellence Base Financing Program of Conicyt. A.N is supported by ANID-Subdirección de Capital Humano/Doctorado Nacional/2025-21253071. M.M. thanks Silvia Trabucco and Elena Poli for very useful discussion. 

\section{Appendix}
\label{sec:appendix}
\begin{figure}[t!]
   \centering
    \includegraphics[width=\columnwidth]{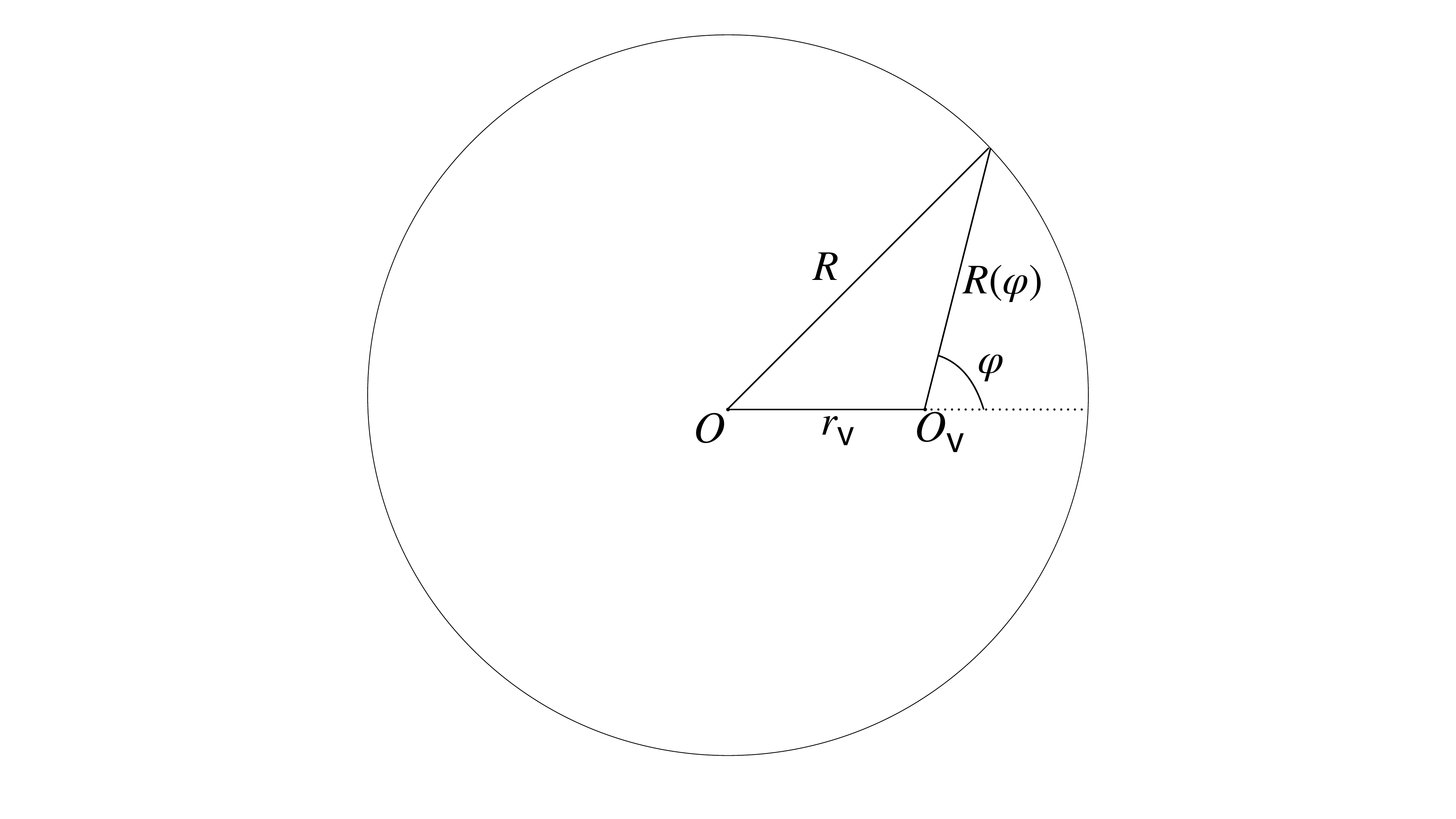}\qquad
            \caption{
            Two-dimensional projection of the cylindrical system  useful for the evaluation of the angular momentum of a noncentral vortex. $O$ coincides with the axis of the cylinder while  $O_\text{v}$ indicates the vortex radial position. Both the cylinder axis and the vortex are orthogonal to the figure.  }
    \label{fig:vortex_radius}
\end{figure}

Here we give more details on the calculation of the single noncentral vortex angular momentum reported in Fig.\,\ref{fig:Ltot_g_l}. In analogy with Eq.\,\eqref{eq:nIn}, the number density is 
\be
n_I = 16 K \mu_I \frac{|\bm r - \bm r_\text{v}|^{2 n}}{(1+|\bm r - \bm r_\text{v}|^{2 n})^2}\,,
\ee
where $\bm r_\text{v}$ is the vortex radial position from the axis of the cylinder.  It is  convenient to take a reference frame centered at the vortex core, $O_\text{v}$, so that
\be
n_I = 16 K \mu_I \frac{r^{2 n}}{(1+r^{2 n})^2}\,,
\ee
 while the axis of the cylinder passes through  $O$, see Fig.\,\ref{fig:vortex_radius}. 
The angular momentum with respect to the axis of the cylinder  is then given by 
\be
L_z= \int d^3 r (\bm r' \times \bm T)_z\,, 
\ee
where  $\bm r' = \bm r - \bm r_\text{v}$ and $\bm T = (0, T_{0\varphi})$, because there is no momentum flux emanating from the vortex center.
 We obtain that
\be
L_z= L_z^1+L_z^2\,,
\ee
where
\be
L_z^1= n\int d^3 r\, n_I\,,
\ee
and
\be
L_z^2= -  \int d^3 r (\bm r_\text{v} \times \bm T)_z\,.
\ee
Both integrals can be  evaluated as follows:
\be
L_z^1= n\int d^3 r  n_I = n \ell_z \int_{0}^{2 \pi} d \varphi   \int_0^{R (\varphi)} dr r\, n_I\,,
\ee
and
\begin{align}
L_z^2&= r_1\int d^3 r  T_{0 \varphi}\cos{\varphi} \nonumber\\&= n \ell_z r_1 \int_{0}^{2 \pi} d \varphi  \cos{\varphi} \int_0^{R (\varphi)} dr\, n_I\,,
\end{align}
where
\be
R (\varphi) = -r_\text{v} \cos{\varphi} + \sqrt{R^2 -r_\text{v}^2 \sin^2{\varphi} }\,,
\ee
is the distance of the boundary from the vortex center. From   Fig.\,\ref{fig:vortex_radius} we have that
\be
R^2 = r_\text{v}^2 + R(\varphi)^2 + 2 r_\text{v}  R(\varphi) \cos{\varphi}\,,
\ee
and thus the above equation follows. The  integrals can be, in part, analytically solved  using the change of variable in Eq.\,\eqref{eq:change}.

%

\end{document}